\documentclass[doc,apacite]{apa6}\usepackage[]{graphicx}\usepackage[]{color}
\makeatletter
\def\maxwidth{ %
  \ifdim\Gin@nat@width>\linewidth
    \linewidth
  \else
    \Gin@nat@width
  \fi
}
\makeatother

\definecolor{fgcolor}{rgb}{0.345, 0.345, 0.345}
\newcommand{\hlnum}[1]{\textcolor[rgb]{0.686,0.059,0.569}{#1}}%
\newcommand{\hlstr}[1]{\textcolor[rgb]{0.192,0.494,0.8}{#1}}%
\newcommand{\hlcom}[1]{\textcolor[rgb]{0.678,0.584,0.686}{\textit{#1}}}%
\newcommand{\hlopt}[1]{\textcolor[rgb]{0,0,0}{#1}}%
\newcommand{\hlstd}[1]{\textcolor[rgb]{0.345,0.345,0.345}{#1}}%
\newcommand{\hlkwa}[1]{\textcolor[rgb]{0.161,0.373,0.58}{\textbf{#1}}}%
\newcommand{\hlkwb}[1]{\textcolor[rgb]{0.69,0.353,0.396}{#1}}%
\newcommand{\hlkwc}[1]{\textcolor[rgb]{0.333,0.667,0.333}{#1}}%
\newcommand{\hlkwd}[1]{\textcolor[rgb]{0.737,0.353,0.396}{\textbf{#1}}}%

\usepackage{framed}
\makeatletter
\newenvironment{kframe}{%
 \def\at@end@of@kframe{}%
 \ifinner\ifhmode%
  \def\at@end@of@kframe{\end{minipage}}%
  \begin{minipage}{\columnwidth}%
 \fi\fi%
 \def\FrameCommand##1{\hskip\@totalleftmargin \hskip-\fboxsep
 \colorbox{shadecolor}{##1}\hskip-\fboxsep
     \hskip-\linewidth \hskip-\@totalleftmargin \hskip\columnwidth}%
 \MakeFramed {\advance\hsize-\width
   \@totalleftmargin\z@ \linewidth\hsize
   \@setminipage}}%
 {\par\unskip\endMakeFramed%
 \at@end@of@kframe}
\makeatother

\definecolor{shadecolor}{rgb}{.97, .97, .97}
\definecolor{messagecolor}{rgb}{0, 0, 0}
\definecolor{warningcolor}{rgb}{1, 0, 1}
\definecolor{errorcolor}{rgb}{1, 0, 0}
\newenvironment{knitrout}{}{} 

\usepackage{alltt} 
\usepackage{epsfig,amsmath,alltt,setspace,bm,url,subcaption, amsfonts,amsthm}
\usepackage[english]{babel}
\usepackage{bigints}

\title{Computation and application of generalized linear mixed model derivatives using \pkg{lme4}}
\fourauthors{Ting Wang}{Benjamin Graves}{Yves Rosseel}{Edgar C.\ Merkle}
\fouraffiliations{American Board of Family Medicine}{University of Missouri}{Ghent University}{University of Missouri}

\abstract{Maximum likelihood estimation of generalized linear mixed models (GLMMs) is difficult due to marginalization of the random effects. Derivative computations of a fitted GLMM's likelihood is also difficult, especially because the derivatives are not by-products of popular estimation algorithms. In this paper, we first describe theoretical results related to GLMM derivatives along with a quadrature method to efficiently compute the derivatives, focusing on fitted \pkg{lme4} models with a single clustering variable. We describe how psychometric results related to item response models are helpful for obtaining the derivatives, as well as for verifying the derivatives' accuracies. We then provide a tutorial on the many possible uses of these derivatives, including robust standard errors, score tests of fixed effect parameters, and likelihood ratio tests of non-nested models. The derivative computation methods and applications described in the paper are all available in easily-obtained \proglang{R} packages.}

\authornote{This work was supported by NSF grant 1460719.
  Correspondence to Edgar Merkle, University of Missouri.
  Email: merklee@missouri.edu.}
\shorttitle{GLMM Derivatives}

\let\proglang=\textsf
\let\pkg=\emph
\let\code=\texttt

\IfFileExists{upquote.sty}{\usepackage{upquote}}{}
\begin{document}
\maketitle

\section{Introduction}
Maximum likelihood estimation of generalized linear mixed models \cite<GLMMs; e.g.,>{stroup12} is notoriously complicated due to the fact that random effects are integrated out of the model likelihood. In general, the integrals cannot be solved analytically, which means that we must use numerical methods to approximate the integrals. Along with model estimation, these issues make it difficult to apply other statistical methods to estimated GLMMs, because the required pieces of the estimated model are not generally available. For example, consider the computation of ``robust'' (Huber-White) standard errors \cite<e.g.,>{white80, huber67}, as applied to GLMM. In addition to the model's maximum likelihood estimates, we require first and second partial derivatives of the model's likelihood function. These derivatives also require integral approximations, which do not necessarily arise as by-products of the model estimation algorithm.

Of primary importance for this paper, the partial derivatives do not arise as by-products of model estimation via the \pkg{lme4} package \cite{lme4}. This package uses a penalized, iteratively re-weighted least squares (PIRLS) algorithm that indirectly maximizes the marginal likelihood by optimizing a second function that involves conditional random effects \cite<conditional on random effect (co-)variances;>{bates2021}. Although this conditional approach bypasses the difficult integration, it also loses the ability to produce the likelihood derivatives of interest. This makes it difficult to apply many relevant methods that are already implemented within the \proglang{R} ecosystem, including sandwich estimators from package \pkg{sandwich} \cite{sand1, sandwichb, sandwichc}, score-based tests from \pkg{strucchange} \cite{strucchange}, model-based recursive partitioning from \pkg{partykit} \cite{party}, and Vuong tests from \pkg{nonnest2} \cite{nonnest2}. These packages all rely on partial derivatives of the model likelihood function (evaluated at the maximum likelihood estimates, {\em after} model estimation), which to date have not been available for GLMMs estimated by \pkg{lme4}.
So the overall goal of this work is to connect existing statistical 
methods with GLMMs estimated by \pkg{lme4}. The paper's contributions towards this goal include (i) theoretical background on GLMM derivatives, as well as a quadrature method that capitalizes on the fact that we are dealing with estimated models; (ii) a general-purpose implementation of the methods via the \pkg{merDeriv} package; and (iii) a tutorial on how these derivatives can be used in applied research settings, including a variety of \proglang{R} examples.

Our derivations are informed by previous results from both statistics and psychometrics, which include diverse motivations for the GLMM. In particular, the statistics community often views the GLMM as an extension of the linear mixed model, whereas the psychometrics community additionally considers connections between the GLMM and item response theory (IRT) models \cite<e.g.,>{de2011,dorbat07}. The latter connections are seldom
noticed in the statistics literature, though \citeA{skrrab04} is noteworthy in that LMMs, GLMMs, and IRT models are included within a larger latent variable framework. We describe below how IRT results can help us obtain derivatives of the GLMM likelihood function with respect to both fixed parameters and random effect hyperparameters (e.g., random effect variances) after model estimation. 

In the following sections, we first fix notation and define the GLMM. We then present theoretical results related to derivatives of the GLMM likelihood function, including a quadrature method that can be applied to estimated models. Next, we provide a tutorial on the application of these results to GLMMs estimated via \pkg{lme4}. This is accomplished with the help of \proglang{R} package \pkg{merDeriv} \cite{merDeriv}, which implements the methods described here, combined with other packages like \pkg{mirt} \cite{mirt}, \pkg{sandwich}, \pkg{nonnest2}, and \pkg{strucchange}. Finally, we discuss potential future extensions of our work.

\section{Theoretical Background}
Our presentation of the GLMM follows the \pkg{lme4} framework of \citeA{lme4}, which facilitates the \proglang{R} applications presented later. This framework encompasses a variety of GLMMs from the exponential family, with binomial models being especially popular. The framework does not allow for products between free parameters and random effects, which becomes important when we discuss relationships between GLMMs and IRT models below \cite<also see>{de2011,dorbat07}.

\subsection{Model and Notation}
 Let $\bm{y}_i$ be a vector containing the response variable for the $i$th cluster, each entry of which is assumed to follow a specific probability distribution (e.g., binomial or Poisson).  The sample size of cluster $i$ is denoted as $n_i$, so the total sample size across all $I$ clusters is given as $N = \sum_{i=1}^{I}n_i$.  Let $\bm{X}_i$ be the $n_i \times p$ design matrix corresponding to fixed effects for cluster $i$; $\bm{\beta}$ is the fixed effect vector of length $p$; $\bm{Z}_i$ is the $n_i \times q$ design matrix corresponding to random effects for cluster $i$;
and $\bm{u}_i$ is the random effect vector of length $q$.
Then the model can be written as
\begin{align}
    \label{eq:glmmmean}
    E(\bm y_i| \bm u_i, \bm \Lambda_{\bm \theta}) &= \bm \mu_i | \bm \Lambda_{\bm \theta}, \bm u_i \\
      \label{eq:glmmlink}
    \bm \mu_i &= g^{-1}(\bm \eta_i | \bm \Lambda_{\bm \theta}, \bm u_i) \\
      \label{eq:glmmlinear}
    \bm \eta_i &= \bm X_i \bm \beta + \bm Z_i\bm b_i\\
    \label{eq:glmmb}
    \bm b_i &= \bm \Lambda_{\bm \theta} \bm u_i\\
      \label{eq:glmmran}
    \bm u_i &\sim N(\bm 0, \bm I_{q}).
\end{align}
The above equations express the idea that the
bounded support of the
expected value of $\bm y_i$ can be transformed to an unbounded
support of the linear combination $\bm X_i \bm \beta + \bm Z_i\bm b_i$
through the link function $g()$.  The random effects are in $\bm b_i$, which equals
 $\bm \Lambda_{\bm \theta} \bm u_i$. The vector $\bm u_i$ follows the standard normal
distribution $N(\bm 0, \bm I_{q})$, with $\bm \Lambda_{\bm \theta}$ being the
relative covariance factor, which can be seen as the Cholesky decomposition of
the usual random effect covariance matrix $\bm{G}$.
Reparameterizing $\bm b_i$ as the product
of the relative covariance factor and standard normal distribution makes it easier to compare GLMM to IRT.  We provide further discussion of this comparison in the next section.

Following the above notation, the model's log-likelihood (marginal over random effects) can be expressed as 
\begin{equation}
  \label{eq:ell}
  \ell = \displaystyle \sum_{i=1}^{I} \ell_i = \displaystyle \sum_{i=1}^I \log \int f_{\bm y_i|\bm u_i}(\bm y_i|\bm u_i)f_{\bm u_i}(\bm u_i)d\bm u_i,
\end{equation}
where $I$ represents the number of total clusters. We further define the following ``across-cluster'' matrices:
\begin{eqnarray}
  \label{eq:redefine1}
  \bm y &=& \{\bm y_1, \bm y_2, \ldots, \bm y_i, \ldots, \bm y_{I} \}\\
  \label{eq:redefine2}
  \bm X &=& \{\bm X_1, \bm X_2, \ldots, \bm X_i, \ldots, \bm X_{I} \}\\
  \label{eq:redefine3}
  \bm Z &=& \{\bm Z_1, \bm Z_2, \ldots, \bm Z_i, \ldots, \bm Z_{I} \}\\
  \label{eq:redefine4}
  \bm b &=& \{\bm b_1, \bm b_2, \ldots, \bm b_i, \ldots, \bm b_{I} \}.
\end{eqnarray}

\subsection{GLMM scores}
One of the most popular IRT models is the two-parameter logistic model \cite<e.g.,>{embrei00,lornov68}, which can be viewed as a binomial GLMM with logit link function.  Consider
an IRT model parameterized as $\text{logit}^{-1}(p_{ij}) = \alpha_j\theta_i - \beta_j$,
with each item $j$'s difficulty described by $\beta_j$ and
discrimination described by $\alpha_j$. The alternative parameterization as $\alpha_j(\theta_i - \beta_j)$ is also applicable, but less convenient for comparison. In the former parameterization, the IRT $\beta_j$ parameters are similar to the negative of
the GLMM fixed parameter $\bm \beta$. The IRT $\alpha_j$ parameters
are then similar to the relative covariance factor in the GLMM, with the \pkg{lme4} package requiring the covariance factor to be equal for all items. This means that we cannot fit a 2PL model in \pkg{lme4}, though other GLMM software such as SAS PROC NLMIXED may allow for 2PL estimation.

In the context of IRT, 
\citeA{glas92,glas98,glas99} utilized an identity from \citeA{lou82} to obtain first derivatives of the marginal log-likelihood (marginal over person parameters $\theta_i$). This identity can be used to show that the first derivative of the marginal log-likelihood with respect to difficulty and discrimination
parameters equals an expected value involving first derivatives of the conditional likelihood (conditioned on person proficiency). That is, we can obtain derivatives of the marginal likelihood by taking an expected value that involves the conditional likelihood.

The same idea can be applied to
GLMM \cite{mcc01}, where conditioning on person proficiency is replaced with conditioning on random effects.  In the
next sections, we will formalize these GLMM score derivations.  Please note
that, throughout this paper, \emph{scores} refer to first derivatives of the
clusterwise log-likelihood function with respect to some model parameters. They are different from
\emph{factor scores} and from \emph{scoring} in psychometrics, which involve prediction of a model's random parameters.

\subsubsection{Fixed effect scores}
Drawing on derivations by Glas as well as by \citeA{mcc01}, the GLMM score with
respect to the fixed effect
parameter $\bm \beta$ can be expressed in the following form:
\begin{equation}
  \label{eq:scorebeta}
  \frac{\partial \ell_i}{\partial \bm \beta} = \frac{\displaystyle \int
    \frac{\partial
    \log f_{\bm y_i|\bm u_i}(\bm y_i|\bm u_i)}{\partial \bm \beta}
  f_{\bm y_i|\bm u_i}(\bm y_i|\bm u_i)f_{\bm u_i}(\bm u_i)d\bm u_i}{f_{\bm y_i} (\bm y_i)},
\end{equation}
where $f_{\bm y_i} (\bm y_i)= \displaystyle \int f_{\bm y_i|\bm u_i}
(\bm y_i|\bm u_i)f_{\bm u_i}(\bm u_i)d\bm u_i$.

The first term in the numerator of Equation~\eqref{eq:scorebeta} can be seen
as the score of a Generalized Linear Model (GLM), which can be expressed in matrix form as
\begin{equation}
  \label{eq:glmscore}
  \frac{\partial \log f_{\bm y_i|\bm u_i}(\bm y_i|\bm u_i)}{\partial \bm \beta} = \bm X_{i}^{T}\bm
  D_{i}^{-1} \bm V_{i}^{-1}(\bm y_{i} - \bm \mu_{i}),
\end{equation}
where $\bm D_i$ and $\bm V_i$ are $n_i \times n_i$ diagonal matrices with
diagonal entries as
$\frac{\partial (\eta_t|u_t)}{\partial (\mu_t|u_t)}$ and
$a(\phi_t)\text{Var}(\mu_t|u_t)$, respectively. The $t$ subscript indexes
an observation within cluster $i$, $1, 2, ..., n_i$.
Further, the $a(\phi_t)$ function is unique to each distribution from the exponential family.
For example, $a(\phi_t) = 1$ for the binomial distribution and
for the Poisson distribution. The value of $a(\phi_t)$ for
other exponential family distributions can be found in, e.g., \citeA{mccnel89}.
Many of the relevant derivations are also supplied by the \proglang{R} \code{family()} function.
Note that, if we
use the canonical link function,
$\frac{\partial (\eta_t|u_t)}{\partial (\mu_t|u_t)}$
and $\text{Var}(\mu_t|u_t)$
will cancel out.  This feature creates a shortcut for distributions using
the canonical link.

The second term in the numerator of Equation~\eqref{eq:scorebeta} is the
distribution of the GLM given $\bm u_i$.  We use the following matrix form to
express all distributions belonging to the exponential family:
\begin{equation}
  \label{eq:glm}
   f_{\bm y_i|\bm u_i}(\bm y_i|\bm u_i) = \text{exp}\left ( \bm y_{i}^{T} \bm A_i \bm \kappa_i -
   \bm 1^{T} \bm A_i h(\bm \kappa_i) + c(\bm y_i, \bm \psi_i)\right ),
\end{equation}
where $\bm A_i$ is a $n_i \times n_i$ diagonal matrix with diagonal element as
$\frac{1}{a(\phi_t)}$; $\bm \kappa_i$ is the vector of canonical parameters;
$\bm 1$ is a
$n_i \times 1$ vector with each entry as $1$;
$h(\bm \kappa_i)$ is an $n_i \times 1$ vector defined by
applying the distribution-specific function $h()$ to each element
of $\bm \kappa_i$;
and $c(\bm y_i, \bm \psi_i)$ is an $n_i \times 1$ vector of remaining terms not
depending on $\bm \kappa_i$, with $\bm \psi_i$ containing scale parameters.
For exponential distributions, these terms can also be found in
\citeA{mccnel89} or in the \proglang{R} \code{family()} functions.

The above results based on generalized linear models are straightforward, while the difficulty
involves the integration over $\bm u$. In the same spirit, the denominator can be
viewed as the integration of the GLM distribution over the random variable $\bm u$. Both integrals have no closed form for GLMMs. We discuss use of quadrature to approximate the integrals below, after describing derivatives of random effect hyperparameters.

\subsubsection{Random effect hyperparameter scores}
Following the same type of derivation, the scores w.r.t. the random effect
hyperparameters can be seen as the scores w.r.t. parameters in the $\bm \Lambda_{\bm \theta}$
matrix. The derivation can thus be expressed as:
\begin{equation}
  \label{eq:scorebeta2}
  \frac{\partial \ell_i}{\partial \bm \Lambda_{\bm \theta}} =
  \frac{\displaystyle \int
    \frac{\partial
    \log f_{\bm y_i|\bm u_i}(\bm y_i|\bm u_i)}{\partial \bm \Lambda_{\bm \theta}}
  f_{\bm y_i|\bm u_i}(\bm y_i|\bm u_i)f_{\bm u_i}(\bm u_i)d\bm u_i}{f_{\bm y_i} (\bm y_i)},
\end{equation}
 where $\frac{\partial \log f_{\bm y_i|\bm u_i}(\bm y_i|\bm u_i)}{\partial \bm
 \Lambda_{\bm \theta}}$ equals $\bm u_{i}^{T}\frac{\partial \bm
 \Lambda_{\bm \theta}}{\partial \theta}\bm Z_i^{T}(\bm y_i- \bm \mu_i)$, with
 $\frac{\partial \bm \Lambda_{\bm \theta}}{\partial \theta}$ as a
 matrix composed of
 1s (corresponding to a particular random effect hyperparameter $\theta$)
 and 0s (not corresponding to a particular random effect hyperparameter $\theta$).
 This derivation is similar to the score derivation for the IRT discrimination
 parameter.  An equivalent approach is to rearrange terms using the trace operator \cite<e.g.,>{peter08}, which
 results in the expression $\text{Tr}\left ((\bm Z_i^{T}(\bm y_i - \bm \mu_i) \bm u_i^{T})^{T}
 \frac{\partial \bm \Lambda_{\bm \theta}}{\partial \theta}\right )$.

 \subsubsection{Reparameterization}
 As mentioned above, $\bm \Lambda_{\bm \theta}$ is a Cholesky decomposition
 of the usual variance covariance matrix $\bm G$, so our derivations are taken with respect to the Cholesky decomposition.  In order to obtain the scores with respect to the variance-covariance
 parameters contained in $\bm G$, we utilize the chain rule:
 \begin{eqnarray}
   \label{eq:chain1}
    \frac{\partial \ell}{\partial \bm G} &=&  \frac{\partial
      \ell}{\partial \bm \Lambda_{\bm \theta}} \frac{\partial
      \bm \Lambda_{\bm \theta}}{\partial \bm G}\\
    &=& \frac{\partial \ell}{\partial \bm \Lambda_{\bm \theta}}
    \left\{\frac{\partial \bm \Lambda_{\bm \theta} }{\partial (\bm \Lambda_{\bm \theta} \bm \Lambda_{\bm \theta}^{T})}
    \right\}\\
     &=& \frac{\partial \ell}{\partial \bm \Lambda_{\bm \theta}}
    \left\{\frac{\partial (\bm \Lambda_{\bm \theta} \bm \Lambda_{\bm \theta}^{T})}{\partial \bm \Lambda_{\bm \theta}}\right\}^{-1}.
 \end{eqnarray}
 For the entry in row $i$ and column $j$ of $\bm{\Lambda}_{\bm{\theta}}$, we have that
\begin{equation}
  \frac{\partial (\bm \Lambda_{\bm \theta} \bm \Lambda_{\bm \theta}^{T})}{\partial \bm \Lambda_{\bm{\theta}ij}} = \bm{\Lambda}_{\bm{\theta}} \bm{J}_{ji} + \bm{J}_{ij} \bm{\Lambda}_{\bm{\theta}}^{T},
\end{equation}
where $\bm{J}_{ij}$ is a matrix with entry $(i,j)$ equal to 1 and 0 elsewhere. The derivatives with respect to all unique, nonzero entries of $\bm{\Lambda}_{\bm{\theta}}$ can be computed in this manner to obtain the desired scores.

As an alternative to variances and covariances, users may wish to parameterize the model via standard deviations and correlations. The scores with respect to standard deviations and correlations can
be obtained by applying another chain rule to the above scores that are taken with respect to $\bm G$.  For example, assume a GLMM with two correlated random effects. In the variance-covariance parameterization, we would have parameters $\sigma_{0}^2$, $\sigma_{1}^2$, and $\sigma_{01}$, while, in the standard deviation-correlation parameterization, we would have parameters $\sigma_{0}$, $\sigma_{1}$, and $\rho$. Derivatives for the latter parameterization are:
\begin{eqnarray}
  \label{eq:scoreglmmchain}
  \frac{\partial \ell}{\sigma_{0}} &=& \frac{\partial \ell}{\partial \sigma_{0}^2}\frac{\partial \sigma_{0}^2}{\partial \sigma_{0}} \\
  &=& \frac{\partial \ell}{\partial \sigma_{0}^2} (2\sigma_{0})\\
  \frac{\partial \ell}{\partial \sigma_{1}} &=& \frac{\partial \ell}{\partial \sigma_{1}^2}\frac{\partial \sigma_{1}^2}{\partial \sigma_{1}} \\
  &=& \frac{\partial \ell}{\partial \sigma_{1}^2} (2\sigma_{1})\\
    \frac{\partial \ell}{\partial \rho} &=& \frac{\partial \ell}{\partial \rho \sigma_{0}\sigma_{1}}\frac{\partial \rho \sigma_{0} \sigma_{1}}{\partial \rho} \\
  &=& \frac{\partial \ell}{\partial \sigma_{01}} (\sigma_{0}\sigma_{1}).
\end{eqnarray}

\subsubsection{Quadrature}
All the derivatives above involve integrals that marginalize over the model random effects $\bm{u}$. These integrals do not have closed forms, requiring numerical methods for approximation. The method implemented in \proglang{R} package \pkg{merDeriv} is a simplified version of multivariate adaptive Gauss-Hermite quadrature \cite{liupie94,naysmi82}, with the simplifications being based on the fact that we are computing derivatives {\em after} model estimation. This means that we already have information about posterior modes and variances of random effects from \pkg{lme4}, and we can make use of this information in place of the ``adaptive'' part of the algorithm. \citeA{merfur19} recently used a similar method to compute marginal versions of Bayesian information criteria (see especially their Appendix C), with that method being based on earlier methods described by \citeA{pinbat95} and \citeA{rab05}. While it would be possible to simply use a traditional adaptive quadrature method here, we would have to use it separately for each case in the data (because we seek to compute casewise derivatives). This would be much slower and infeasible for many datasets, as compared to our quadrature method described here.

Focusing on the GLMM framework, the integrals from Equations~\eqref{eq:scorebeta} and~\eqref{eq:scorebeta2} are both of the form
\begin{equation}
\displaystyle \int g(\bm y | \bm u, \bm{\omega}) f_{\bm y|\bm u, \bm{\omega}} (\bm y|\bm u, \bm{\omega})f_{\bm u | \bm{\omega}}(\bm u | \bm{\omega})d\bm u,
\end{equation}
where $g()$ differs depending on the integral, and $\bm{\omega}$ is a vector of model parameters excluding the random effects $\bm{u}$. This conditioning on $\bm{\omega}$ is implicit in earlier sections but was excluded to simplify notation.

For a single clustering variable with $I$ levels, the clusters $i$ are independent. Therefore, the above equation can be written as
\begin{equation}
\displaystyle \prod_{i=1}^I \displaystyle \int g(\bm{y}_i | \bm{u}_i, \bm{\omega}) f_{\bm{y}_i|\bm{u}_i, \bm{\omega}} (\bm{y}_i |\bm{u}_i, \bm{\omega})f_{\bm{u}_i | \bm{\omega}}(\bm{u}_i | \bm{\omega})d\bm{u}_i.
\end{equation}
To compute scores, we are interested in the elements of the above product: the integral for each cluster $i$. For $M$ quadrature points, we use Gauss-Hermite quadrature to approximate the integral for cluster $i$ by:
\begin{equation}
\displaystyle \sum_{m=1}^M w^\ast_{im} g(\bm{y}_i | \bm{a}^\ast_{im}, \bm{\omega}) f_{\bm{y}_i|\bm{u}_i, \bm{\omega}} (\bm{y}_i |\bm{a}^\ast_{im}, \bm{\omega}).
\end{equation}
That is, the integral is approximated by a weighted sum of function evaluations, where the functions are evaluated at different random effect values represented by $\bm{a}^\ast_{im}$, $m=1,\ldots,M$. For a random effect of dimension $d$, the quadrature locations and weights are computed by
\begin{align}
  \bm{a}^\ast_{im} &= \tilde{\bm{b}}_i + \tilde{\bm{C}}_i \times \bm{a}_m \\
  w^\ast_{im} &= w_m \times (2\pi)^{d/2} \times \det{(\tilde{\bm{C}}_i)} \times \exp{(0.5 \times \bm{a}_m \bm{a}_m^\prime)} \times \bm{\phi}(\bm{a}^\ast_{im} | \bm{0}, \hat{\bm{G}})
\end{align}
where $\tilde{\bm{b}}_i$ are the posterior modes of random effects for cluster $i$, $\tilde{\bm{C}}_i$ is the Cholesky factor of the {\em conditional} covariance matrix of the random effects for cluster $i$ (obtained from the \pkg{lme4} function \code{ranef()}), $\bm{\phi}()$ is the normal density function, and $\hat{\bm{G}}$ is the estimated covariance matrix of the random effects (obtained from the \pkg{lme4} function \code{VarCorr()}). Finally, $\bm{a}_m$ and $w_m$ are the usual Gauss-Hermite locations and weights, respectively.

\subsubsection{Second derivatives}
While we have focused on first derivatives, the \citeA{lou82} identity can also aid in computation of second derivatives, leading to the model Hessian and information matrix. We do not present the equations here because, for models estimated via \code{glmer()} (but not \code{lmer()}), a Hessian is already computed and stored in the resulting model object (specifically in the \code{optinfo} slot). According to the \pkg{lme4} documentation, this Hessian is computed using a finite difference approach. The \pkg{merDeriv} package provides a convenience function to access this Hessian, and we use it in our applications later.

By default, the \pkg{lme4} Hessian is parameterized via the Cholesky decomposition of random effects.
The Hessian based on the standard deviation/correlation parameterization can alternatively be
obtained via the \code{devfun2()} function in \pkg{lme4}, which uses the profile likelihood. The Hessian for the variance/covariance parameterization is then related to the latter option, through the chain rule mentioned earlier. The \pkg{merDeriv} package incorporates these computations and enables researchers to request the parameterization of interest via the \code{ranpar} argument (taking possible values of \code{"var"}, \code{"sd"}, or \code{"theta"}). This is illustrated in the tutorials below.

\section{Tutorial on the Derivatives' Uses in \proglang{R}}
We now provide a tutorial on \proglang{R} package \pkg{merDeriv}, which can carry out the computations described above and which can be used to solve applied problems. As we go, we provide snippets of code that illustrate how \pkg{merDeriv} interacts with other packages, which readers can adapt to other models and datasets. We first provide some evidence that \pkg{merDeriv} operates in the manner expected, by comparing a Rasch model estimated via \pkg{lme4} to a Rasch model estimate via \pkg{mirt} \cite{mirt}. We then consider a variety of other applications.

\subsection{Verifying the Computations}
Before using the scores from \pkg{merDeriv} in GLMM applications, we use the relationship between GLMM and IRT to verify the correctness of the quadrature implementation. We specifically compare the score computations to those of package \pkg{mirt} \cite{mirt}, which estimates many types of item response models. We make use of the fact that the Rasch model can be estimated as a generalized linear mixed model, which was illustrated by \citeA{de2011}. We also make use of the fact that \pkg{mirt} has its own, independent quadrature method for score computation, which was used by \citeA{schcha19} to apply Vuong tests to item response models.

\subsubsection{Method}
For comparing the two score computation algorithms, we use the LSAT7
data \cite{boclie70} included with \pkg{mirt}. This dataset includes
the item responses (correct/incorrect) of 1,000 individuals across 5
items of the LSAT.

The code in Figure~\ref{fig:cc1} shows how a Rasch model can be fit
to the data using both \pkg{mirt} and \pkg{lme4}.
For \pkg{mirt}, we require the LSAT7 data to be arranged in wide format, where
each row is a person and each column is an item. If we then rearrange the
data to be in long format, as shown in Figure~\ref{fig:cc1}, we can fit the Rasch model via \pkg{lme4}. 
We use the \code{nAGQ} argument to employ adaptive quadrature during \pkg{lme4} model estimation, avoiding the \code{glmer()} default, \code{nAGQ=1}, which uses the Laplace approximation. The quadrature leads to a more accurate approximation of the model log-likelihood, which in turn leads to maximum likelihood estimates that tend to be closer to the true maximum of the likelihood. The \pkg{mirt} package employs a fixed quadrature method with 61 quadrature points.

\begin{figure}
\caption{Code to fit Rasch models using \pkg{mirt} and \pkg{lme4}, then calculate scores.}
\label{fig:cc1}
\begin{knitrout}
\definecolor{shadecolor}{rgb}{0.969, 0.969, 0.969}\color{fgcolor}\begin{kframe}
\begin{alltt}
\hlcom{## mirt:}
\hlkwd{library}\hlstd{(}\hlstr{"mirt"}\hlstd{)}
\hlstd{ls7} \hlkwb{<-} \hlkwd{expand.table}\hlstd{(LSAT7)}
\hlstd{mirtmod} \hlkwb{<-} \hlkwd{mirt}\hlstd{(ls7[,}\hlnum{1}\hlopt{:}\hlnum{5}\hlstd{],} \hlnum{1}\hlstd{,} \hlkwc{itemtype} \hlstd{=} \hlstr{"Rasch"}\hlstd{,} \hlkwc{SE} \hlstd{=} \hlnum{TRUE}\hlstd{)}

\hlcom{## reshape data and fit with glmer():}
\hlkwd{library}\hlstd{(}\hlstr{"reshape2"}\hlstd{)}
\hlstd{ls7}\hlopt{$}\hlstd{person} \hlkwb{<-} \hlnum{1}\hlopt{:}\hlkwd{nrow}\hlstd{(ls7)}
\hlstd{ls7long} \hlkwb{<-} \hlkwd{melt}\hlstd{(ls7,} \hlkwc{id} \hlstd{=} \hlstr{"person"}\hlstd{)}
\hlstd{lme4mod} \hlkwb{<-} \hlkwd{glmer}\hlstd{(value} \hlopt{~ -}\hlnum{1} \hlopt{+} \hlstd{variable} \hlopt{+} \hlstd{(}\hlnum{1} \hlopt{|} \hlstd{person),} \hlkwc{family} \hlstd{= binomial,}
                 \hlkwc{data} \hlstd{= ls7long,} \hlkwc{nAGQ} \hlstd{=} \hlnum{5L}\hlstd{)}

\hlcom{## score calculation:}
\hlstd{mirtsc} \hlkwb{<-} \hlkwd{estfun.AllModelClass}\hlstd{(mirtmod)}
\hlstd{lme4sc} \hlkwb{<-} \hlkwd{estfun.glmerMod}\hlstd{(lme4mod,} \hlkwc{ranpar} \hlstd{=} \hlstr{"var"}\hlstd{)}
\end{alltt}
\end{kframe}
\end{knitrout}
\end{figure}

\subsubsection{Results}
As shown at the bottom of Figure~\ref{fig:cc1}, scores for the two models are obtained via their respective
\code{estfun()} functions. The function for \pkg{mirt} models is
included directly within the \pkg{mirt} package, whereas the function
for \pkg{lme4} models is included in \pkg{merDeriv}.
Both functions output a score matrix, where rows index people and columns index model parameters. For the \code{glmer} model, we use the \code{ranpar} argument so that the \pkg{merDeriv} scores involve the variance-covariance parameterization, which matches the \pkg{mirt} output.

\begin{figure}
\begin{center}
\begin{knitrout}
\definecolor{shadecolor}{rgb}{0.969, 0.969, 0.969}\color{fgcolor}
\includegraphics[width=3.5in]{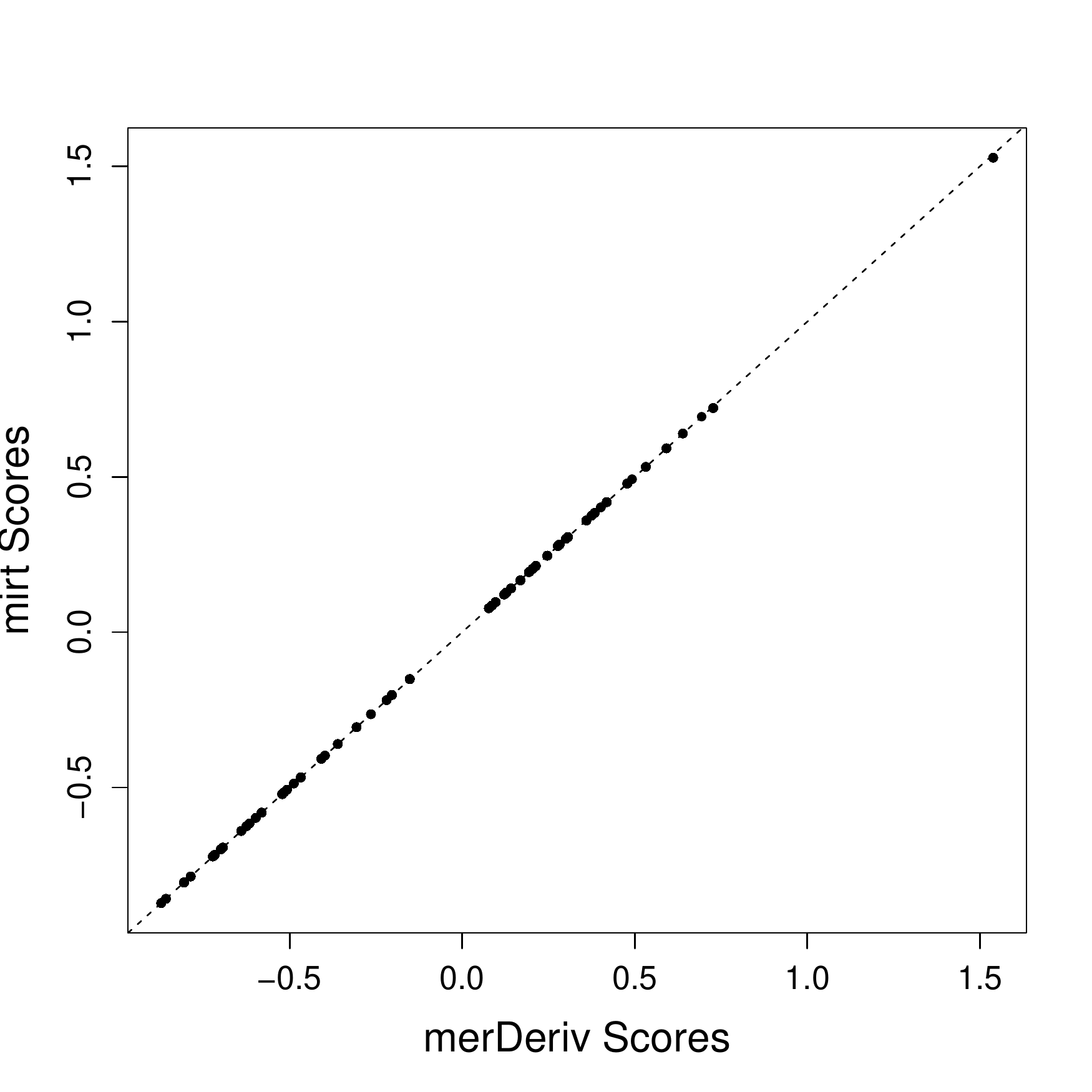} 

\end{knitrout}
\caption{Comparison of Rasch model scores from \pkg{mirt} and from \pkg{merDeriv}.}
\label{fig:verfig}
\end{center}
\end{figure}

In comparing the two sets of scores, we arrive at Figure~\ref{fig:verfig}. The x-axis depicts scores from \pkg{merDeriv}, the y-axis depicts scores from \pkg{mirt}, and each point is a particular score. We see that the values are nearly exactly equal for \pkg{mirt} and for \pkg{merDeriv}, falling directly on the identity line. One can also compare the parameter variance-covariance matrix of \pkg{merDeriv} and of \pkg{mirt}, using the \code{vcov()} method of each package. That comparison, not shown, exhibits agreement similar to the score comparison. These provide evidence that the \pkg{merDeriv} code is performing as expected. Now that we have obtained this evidence, we move on to illustrate practical uses of the scores in GLMM applications.

\begin{table}
  \centering
  \begin{tabular}{lcccccc} \hline
 & $\beta_1$ & $\beta_2$ & $\beta_3$ & $\beta_4$ & $\beta_5$ & $\sigma^2$ \\ \hline
lme4 & 0.1004 & 0.0811 & 0.0913 & 0.0787 & 0.1037 & 0.1300 \\
sandwich & 0.0996 & 0.0814 & 0.0898 & 0.0785 & 0.1058 & 0.1311 \\ \hline

  \end{tabular}
  \caption{Comparison between Rasch model standard errors reported by \pkg{lme4} and robust standard errors reported by \pkg{sandwich}. The $\beta$ columns correspond to item difficulties, while the $\sigma^2$ column corresponds to person (intercept) variance.}
  \label{tab:robse}
\end{table}

\subsection{Huber-White estimator}
Let $\bm{\omega}$ be the model parameter vector, which in a GLMM would contain fixed effect parameters and random effect (co-)variances. Then the Huber-White \cite<e.g.,>{white80, huber67} sandwich estimator of the covariance matrix of $\bm{\omega}$ is
\begin{equation}
  \label{eq:aba}
    \bm V(\hat{\bm \omega}) = (\bm A)^{-1}\bm B(\bm A)^{-1},
\end{equation}
where $\bm{A}$ is the negative expectation of the model Hessian and $\bm{B}$ is the covariance matrix of scores \cite<see>[for further discussion in the context of linear mixed models]{merDeriv}. The score computations described in the previous sections facilitate computation of this $\bm{B}$ matrix. The square root of the diagonal elements of $\bm{V}$ are then typically called ``robust standard errors.''

Robust standard errors are used to address model misspecifications such as unmodeled dependence between observations or deviations from normality.
While random effects are typically used in GLMMs to account for dependence between observations, the Huber-White estimator can be used on top of a GLMM to account for further model misspecifications. Further, \citeA{str20} recently provided evidence that quadrature can lead to downward-biased variance estimates in GLMMs, resulting in inflated Type I error rates. The Huber-White estimator may be considered in light of this result.

We can easily compute Huber-White standard errors using the scores
from the previous section, paired with the \pkg{sandwich} package, as
shown in Figure~\ref{fig:cc4}. In that figure, the \code{bread.glmerMod()} and \code{meat()} functions come from \pkg{merDeriv}, while \code{sandwich()} comes from the \pkg{sandwich} package. Applying this result to the Rasch model estimated in the previous section, we obtain the results in Table~\ref{tab:robse}. For this particular application, the \pkg{lme4} standard errors and \pkg{sandwich} standard errors are virtually equal, likely due to the large sample size (large by GLMM standards, at least).

\begin{figure}
\caption{Example code for calculating Huber-White standard errors.}
\label{fig:cc4}
\begin{knitrout}
\definecolor{shadecolor}{rgb}{0.969, 0.969, 0.969}\color{fgcolor}\begin{kframe}
\begin{alltt}
\hlkwd{library}\hlstd{(}\hlstr{"sandwich"}\hlstd{)}
\hlkwd{sandwich}\hlstd{(lme4mod,} \hlkwc{bread.} \hlstd{= bread.glmerMod,} \hlkwc{meat.} \hlstd{=} \hlkwd{meat}\hlstd{(lme4mod,} \hlkwc{level} \hlstd{=} \hlnum{2}\hlstd{))}
\end{alltt}
\end{kframe}
\end{knitrout}
\end{figure}

\subsection{Score tests}
Researchers have long been familiar with score tests, also known as Lagrange multiplier tests, that can be used as an alternatives to the likelihood ratio test or to the Wald test \cite<e.g.,>{eng84,glas92,glas98,glas99}. In typical score test applications, a constrained model is fit to data, then first derivatives of the likelihood function are used to test whether or not some constraint should be relaxed. In contrast, the likelihood ratio test requires us to estimate two models (a constrained model and an unconstrained model), and the Wald test requires us to estimate only the unconstrained model.

This score test framework has expanded to a class of ``parameter instability'' tests, where we test whether an estimated model's parameters differ with respect to unmodeled auxiliary variables (with different test statistics being used for continuous, ordinal, or discrete auxiliary variables). \citeA{zeihor07} summarized much previous work on this topic, developing a family of score-based tests that can be used within an M-estimation framework (of which maximum likelihood estimation is a special case). They also developed \proglang{R} package \pkg{strucchange} \cite{strucchange}, which can be used to compute the test statistics so long as a model's scores and Hessian are available. The family of score-based tests has subsequently been studied in the context of many specific types of models, including linear mixed models \cite{wanmer20}, structural equation models \cite{merzei13, merfanzei}, and item response models \cite{kom18, strkop15, wanstr18}.
The developments in the current paper make it possible to apply score-based tests to GLMMs, yielding test statistics for GLMMs that have been unavailable up to now.
The score computations described above can be used to construct the cumulative scores, which are further used to compute test statistics \cite{merzei13, merfanzei}.

In this section, we show how scores can be used to test fixed effect parameters that are not directly included in a GLMM model. This is potentially useful in situations where a model with the fixed effect included does not converge, which often happens in applied mixed modeling \cite<see>{barlev13,matkli17}. In these situations, if we can get a model to converge {\em without} some fixed effect of interest, it is possible to apply score-based tests to the fitted model in order to test the omitted fixed effect. While the more popular approach here is to drop random effects (as opposed to fixed effects) from the model, dropping fixed effects may be useful in instances where, e.g., the random effect variances are all large, yet the model still exhibits convergence problems.

\subsubsection{Method}
We use data from 500 respondents on the Nerdy Personality Attributes Scale (NPAS), a personality test designed for personal entertainment on the Open Source Psychometrics Project website \cite{ospp}. The questionnaire consists of 26 items that attempt to define the concept of ``nerdiness''. Responses were originally measured on 5-point Likert scales, but we converted them to binary responses for this example (where 0 corresponds to 3 or less and 1 corresponds to 4 or 5). The items ask about different aspects of nerdiness, including hobbies and interests that are usually associated with nerds, social interactions, personality traits, and academic or intellectual endeavors. The data also include various demographic variables and other personality measures assessing the ``Big Five'' personality factors.

Here, we assess whether item responses vary across extraversion, while also accounting for
inherent item differences (which would be called ``item difficulties''
in an IRT context). The \pkg{lme4} syntax for this model is shown at the top of Figure~\ref{fig:cc5}, where the variable names are generally self-explanatory. Note that inclusion of the interaction term (\code{item*ext}) automatically includes main effects of both item and extraversion, in addition to the interaction. This GLMM can be viewed as a person-by-covariate item response model, falling into the class of explanatory item response models considered by \citeA{debwil04} and \citeA{de2011}.

\begin{figure}
\caption{Models of the NPAS data. The first model has issues with non-convergence, leading us to the simpler, second model. A score test is then used to study the interaction.}
\label{fig:cc5}
\begin{knitrout}
\definecolor{shadecolor}{rgb}{0.969, 0.969, 0.969}\color{fgcolor}\begin{kframe}
\begin{alltt}
\hlcom{## Model that has problems with convergence:}
\hlstd{m1} \hlkwb{<-} \hlkwd{glmer}\hlstd{(answer} \hlopt{~ -}\hlnum{1} \hlopt{+} \hlstd{item}\hlopt{*}\hlstd{ext} \hlopt{+} \hlstd{(}\hlnum{1} \hlopt{|} \hlstd{subject),}
            \hlkwc{data} \hlstd{= npas.sampled,} \hlkwc{family} \hlstd{= binomial)}

\hlcom{## Model with only main effects, which converges:}
\hlstd{m2} \hlkwb{<-} \hlkwd{glmer}\hlstd{(answer} \hlopt{~ -}\hlnum{1} \hlopt{+} \hlstd{item} \hlopt{+} \hlstd{ext} \hlopt{+} \hlstd{(}\hlnum{1} \hlopt{|} \hlstd{subject),}
            \hlkwc{data} \hlstd{= npas.sampled,} \hlkwc{family} \hlstd{= binomial,}
            \hlkwc{control} \hlstd{=} \hlkwd{glmerControl}\hlstd{(}\hlkwc{optimizer}\hlstd{=}\hlstr{'bobyqa'}\hlstd{))}

\hlcom{## Score test:}
\hlstd{ext} \hlkwb{<-} \hlkwd{with}\hlstd{(npas.sampled,} \hlkwd{as.numeric}\hlstd{(}\hlkwd{tapply}\hlstd{(ext, subject, head,} \hlnum{1}\hlstd{)))}

\hlstd{sc1} \hlkwb{<-} \hlkwd{sctest}\hlstd{(m2,} \hlkwc{fit}\hlstd{=}\hlkwa{NULL}\hlstd{,} \hlkwc{scores}\hlstd{=estfun.glmerMod,} \hlkwc{order.by}\hlstd{=ext,}
       \hlkwc{parm}\hlstd{=}\hlnum{1}\hlopt{:}\hlnum{26}\hlstd{,} \hlkwc{functional}\hlstd{=}\hlstr{'maxLMo'}\hlstd{)}
\end{alltt}
\end{kframe}
\end{knitrout}
\end{figure}

\subsubsection{Results}
The first model in Figure~\ref{fig:cc5} did not converge, even after making changes to the optimizer and its settings. 
We could have experimented further, perhaps finding some combination of settings that led to a converged model and that would render the score test unnecessary. But each attempted model estimation took about ten minutes, so we could easily have spent hours tweaking the settings. In contrast, the score test could be immediately applied to a simpler model that converged more easily.

Our simpler model was the second model in Figure~\ref{fig:cc5}, which used the \verb+bobyqa+ optimizer
\cite{powell2009} instead of the default \verb+Nelder_Mead+. In estimating this second model, we capitalize on the fact that score tests require only a ``constrained'' model, which here assumes that responses to items do not vary across levels of extraversion. We can then obtain a score test statistic for the interaction without directly including the interaction in the model.

\begin{figure}
\caption{M-fluctuation test for NPAS data. This graph presents item parameter fluctuation across varying levels of extraversion. Peaks of the graph suggest extraversion cutpoints that isolate individuals with similar item parameters.}
\label{fig:mfluc}
  \begin{center}
\begin{knitrout}
\definecolor{shadecolor}{rgb}{0.969, 0.969, 0.969}\color{fgcolor}
\includegraphics[width=3.5in]{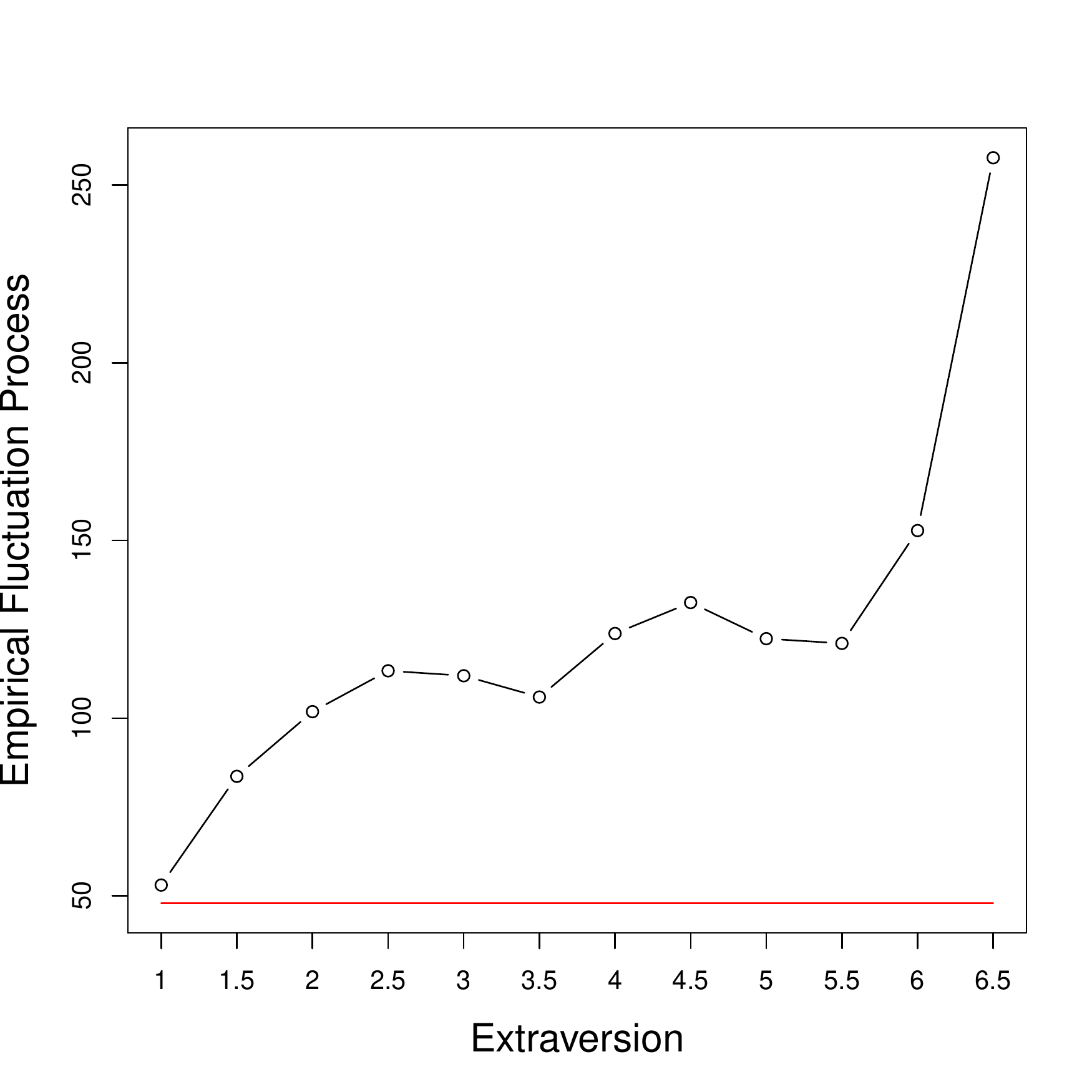} 

\end{knitrout}
\end{center}
\end{figure}

To obtain a test statistic for this interaction, we carry out the score test using the code at the bottom of Figure~\ref{fig:cc5}. This makes use of the \code{sctest()} function found in the \proglang{R} package \pkg{strucchange} as well as \code{estfun.glmerMod()} function found in \pkg{merDeriv}. It simultaneously tests all 26 item parameters for fluctuations with respect to extraversion, which is similar to including an item$\times$extraversion interaction.
Results of this test are visualized in Figure~\ref{fig:mfluc}, which shows how the scores fluctuate across different values of extraversion (x-axis). We can see that there is significant parameter fluctuation in items across levels of extraversion, because the black line goes above the ``critical value'' that is depicted by the red line. The peaks in the black line, around extraversion values of 2.5 and 4.5, suggest cutpoints for subgroups of individuals that exhibit similar item parameters. The test provides information about the nature of the interaction that was not easily obtained by including extraversion in the model, due to model convergence problems.

\subsection{Vuong tests}
Scores also play a role in Vuong tests \cite{vuo89}, which can be used to compare nested and non-nested models to one another. In the nested case, the tests can be viewed as extensions of the traditional likelihood ratio test, which, unlike the traditional likelihood ratio test, make no assumptions about the more complex model being correct. In the non-nested case, the tests provide a formal way of comparing the fits of the two models.
The scores described in this paper can be used in tandem with package \pkg{nonnest2} \cite{nonnest2} to compare GLMMs, providing new capabilities for comparing models with different predictor variables and different random effects.
Specifically, our score computations are used to compute the null distribution of the test statistic, which is a weighted
sum of chi-square distributions. Further descriptions of the tests and applications to psychometric models can be found in \citeA{merkle16} and in \citeA{schcha19}. An illustration involving GLMMs is provided here.

\subsubsection{Method}
The data used for this example comes from the SPISA data set, which can be found in the \proglang{R} package \pkg{psychotree} \cite{strkop15}. The data is a subsample of 1,075 Bavarian university students who took an online, general knowledge quiz called ``Studentenpisa'' administered by a German weekly news magazine \cite{trever10}. The quiz consists of 45 items on 5 topics, and we focus here on a subset of nine questions dealing with natural science. The data set includes several covariates such as age, gender, semester of university enrollment, and elite university status.

Using a similar item response model as in the previous example, we construct two non-nested models with different covariates. These models are based on a common reduced model that only contains item and subject effects. The first model uses age and gender as covariates, while the second model uses semester of university enrollment and whether the student's university has been granted ``elite'' status or not. The code for these models is shown in Figure~\ref{fig:cc11}. Similar to the previous application, the models here did not immediately converge, and we switched optimizers in order to attain convergence. Following model estimation, we obtained scores and compared the two models using a Vuong test computed via the \proglang{R} package \pkg{nonnest2} \cite{nonnest2}.

\begin{figure}
\caption{Code for the non-nested models to be compared using the Vuong test. The first model uses age and gender as potential predictors, while the second model uses number of semesters at the university and elite university status.}
\label{fig:cc11}
\begin{knitrout}
\definecolor{shadecolor}{rgb}{0.969, 0.969, 0.969}\color{fgcolor}\begin{kframe}
\begin{alltt}
\hlstd{mod1} \hlkwb{<-} \hlkwd{glmer}\hlstd{(response} \hlopt{~ -}\hlnum{1} \hlopt{+} \hlstd{item} \hlopt{+} \hlstd{agecent} \hlopt{+} \hlstd{gender} \hlopt{+} \hlstd{(}\hlnum{1} \hlopt{|} \hlstd{pnum),}
              \hlkwc{data} \hlstd{= spisa,} \hlkwc{family} \hlstd{= binomial,}
              \hlkwc{control} \hlstd{=} \hlkwd{glmerControl}\hlstd{(}\hlkwc{optimizer}\hlstd{=}\hlstr{'bobyqa'}\hlstd{))}

\hlstd{mod2} \hlkwb{<-} \hlkwd{glmer}\hlstd{(response} \hlopt{~ -}\hlnum{1} \hlopt{+} \hlstd{item} \hlopt{+} \hlstd{semester} \hlopt{+} \hlstd{elite} \hlopt{+} \hlstd{(}\hlnum{1} \hlopt{|} \hlstd{pnum),}
              \hlkwc{data} \hlstd{= spisa,} \hlkwc{family} \hlstd{= binomial,}
              \hlkwc{control} \hlstd{=} \hlkwd{glmerControl}\hlstd{(}\hlkwc{optimizer}\hlstd{=}\hlstr{'bobyqa'}\hlstd{))}
\end{alltt}
\end{kframe}
\end{knitrout}
\end{figure}

\subsubsection{Results}

The \pkg{nonnest2} code and output for the Vuong test is shown in Figure~\ref{fig:cc12}. First, we create a convenience function, \code{vcg()}, to compute the full parameter covariance matrix (including random effect variances/covariances) for each of the models. This function, along with functions from \pkg{merDeriv} for calculating the likelihoods and scores, is then sent to \code{vuongtest()}.

\begin{figure}
\caption{Code to run Vuong test for comparing two non-nested models. The models are able to be distinguished from each other, but one model does not have better fit over the other.}
\label{fig:cc12}
\begin{knitrout}
\definecolor{shadecolor}{rgb}{0.969, 0.969, 0.969}\color{fgcolor}\begin{kframe}
\begin{alltt}
\hlstd{vcg} \hlkwb{<-} \hlkwa{function}\hlstd{(}\hlkwc{obj}\hlstd{)} \hlkwd{vcov}\hlstd{(obj,} \hlkwc{full} \hlstd{=} \hlnum{TRUE}\hlstd{)}

\hlkwd{vuongtest}\hlstd{(mod1, mod2,} \hlkwc{ll1} \hlstd{= llcont.glmerMod,} \hlkwc{ll2} \hlstd{= llcont.glmerMod,}
          \hlkwc{score1} \hlstd{= estfun.glmerMod,} \hlkwc{score2} \hlstd{= estfun.glmerMod,}
          \hlkwc{vc1} \hlstd{= vcg,} \hlkwc{vc2} \hlstd{= vcg)}
\end{alltt}
\begin{verbatim}

Model 1 
 Class: glmerMod 
 Call: glmer(formula = response ~ -1 + item + agecent + gender + (1 | ...

Model 2 
 Class: glmerMod 
 Call: glmer(formula = response ~ -1 + item + semester + elite + (1 | ...

Variance test 
  H0: Model 1 and Model 2 are indistinguishable 
  H1: Model 1 and Model 2 are distinguishable 
    w2 = 0.033,   p = 6.25e-07

Non-nested likelihood ratio test 
  H0: Model fits are equal for the focal population 
  H1A: Model 1 fits better than Model 2 
    z = -0.356,   p = 0.639
  H1B: Model 2 fits better than Model 1 
    z = -0.356,   p = 0.3611
\end{verbatim}
\end{kframe}
\end{knitrout}
\end{figure}

The output from the function first shows a variance test, which provides information about whether the non-nested models are distinguishable from each other via the observed dataset. From this, we reject the hypothesis that the models are indistinguishable from one another. We then move on to the non-nested likelihood ratio test to examine whether one model fits better than the other. For our example, we conclude that neither model fits better than the other.

Figure~\ref{fig:cc13} shows that the \pkg{nonnest2} functionality can also be used to test nested models, by adding the \code{nested = TRUE} argument. We first fit a simple Rasch model to the data, with this model being nested in the two considered previously. We then compute test statistics comparing this model to the second model from Figure~\ref{fig:cc11}. The two test statistics in the output can each be used to compare the nested models, providing two alternatives to the traditional likelihood ratio test. Here, we conclude that the full model including the ``semester'' and ``elite'' predictors fits better than the simple Rasch model without those predictors.

\begin{figure}
\caption{Code for testing fit of two nested models. The full model has better fit than the reduced model.}
\label{fig:cc13}
\begin{knitrout}
\definecolor{shadecolor}{rgb}{0.969, 0.969, 0.969}\color{fgcolor}\begin{kframe}
\begin{alltt}
\hlstd{mod3} \hlkwb{<-} \hlkwd{glmer}\hlstd{(response} \hlopt{~ -}\hlnum{1} \hlopt{+} \hlstd{item} \hlopt{+} \hlstd{(}\hlnum{1} \hlopt{|} \hlstd{pnum),} \hlkwc{data} \hlstd{= spisa,}
              \hlkwc{family} \hlstd{= binomial,}
              \hlkwc{control} \hlstd{=} \hlkwd{glmerControl}\hlstd{(}\hlkwc{optimizer}\hlstd{=}\hlstr{'bobyqa'}\hlstd{))}

\hlkwd{vuongtest}\hlstd{(mod2, mod3,} \hlkwc{nested} \hlstd{=} \hlnum{TRUE}\hlstd{,}
          \hlkwc{ll1} \hlstd{= llcont.glmerMod,} \hlkwc{ll2} \hlstd{= llcont.glmerMod,}
          \hlkwc{score1} \hlstd{= estfun.glmerMod,} \hlkwc{score2} \hlstd{= estfun.glmerMod,}
          \hlkwc{vc1} \hlstd{= vcg,} \hlkwc{vc2} \hlstd{= vcg)}
\end{alltt}
\begin{verbatim}

Model 1 
 Class: glmerMod 
 Call: glmer(formula = response ~ -1 + item + semester + elite + (1 | ...

Model 2 
 Class: glmerMod 
 Call: glmer(formula = response ~ -1 + item + (1 | pnum), data = spisa, ...

Variance test 
  H0: Model 1 and Model 2 are indistinguishable 
  H1: Model 1 and Model 2 are distinguishable 
    w2 = 0.017,   p = 0.000109

Robust likelihood ratio test of distinguishable models 
  H0: Model 2 fits as well as Model 1 
  H1: Model 1 fits better than Model 2 
    LR = 18.680,   p = 9.08e-05
\end{verbatim}
\end{kframe}
\end{knitrout}
\end{figure}

\subsection{Poisson GLMMs}
Of course, the GLMM framework is not limited solely to binomial models, and our derivations extend to other exponential family models. In this section, we illustrate extensions to the Poisson GLMM using the epilepsy data set \cite{thall1990} found in the package \pkg{brms} \cite{brms}.

\subsubsection{Method}
The data consist of 236 observations of seizure counts from 59 people across 4 time periods. Covariates include study group (treatment vs control), participant age, and a base rate seizure count across 8-weeks (standardized). For our initial model, we predict number of seizures using the patient's base rate (\code{zBase}), treatment group indicator (\code{Trt}), and visit number (\code{visit}). We allow the intercept and \code{visit} slope to vary by participant, with these two random effects being correlated. The \pkg{lme4} code for this model is at the top of Figure~\ref{fig:cc8}.

\begin{figure}
\caption{Code to fit a Poisson GLMM predicting the number of seizures in epileptic patients, then compute robust standard errors and a score test statistic.}
\label{fig:cc8}
\begin{knitrout}
\definecolor{shadecolor}{rgb}{0.969, 0.969, 0.969}\color{fgcolor}\begin{kframe}
\begin{alltt}
\hlcom{## linear effect of visit number:}
\hlstd{epilepsy}\hlopt{$}\hlstd{visit} \hlkwb{<-} \hlkwd{as.numeric}\hlstd{(epilepsy}\hlopt{$}\hlstd{visit)}

\hlcom{## Poisson model:}
\hlstd{poimod} \hlkwb{<-} \hlkwd{glmer}\hlstd{(count} \hlopt{~} \hlstd{zBase} \hlopt{*} \hlstd{Trt} \hlopt{*} \hlstd{visit} \hlopt{+} \hlstd{(visit} \hlopt{|} \hlstd{patient),}
                \hlkwc{data} \hlstd{= epilepsy,} \hlkwc{family} \hlstd{= poisson)}

\hlcom{## Robust standard errors:}
\hlstd{rse} \hlkwb{<-} \hlkwd{sandwich}\hlstd{(poimod,} \hlkwc{bread.} \hlstd{= bread.glmerMod,}
                \hlkwc{meat.} \hlstd{=} \hlkwd{meat}\hlstd{(poimod,} \hlkwc{level} \hlstd{=} \hlnum{2}\hlstd{))}

\hlcom{## Score-based test with 5 quadrature points:}
\hlstd{age} \hlkwb{<-} \hlkwd{with}\hlstd{(epilepsy,} \hlkwd{tapply}\hlstd{(Age, patient, head,} \hlnum{1}\hlstd{))}
\hlstd{efg5} \hlkwb{<-} \hlkwa{function}\hlstd{(}\hlkwc{...}\hlstd{)} \hlkwd{estfun.glmerMod}\hlstd{(...,} \hlkwc{nAGQ} \hlstd{=} \hlnum{5}\hlstd{)}

\hlstd{poisc} \hlkwb{<-} \hlkwd{sctest}\hlstd{(poimod,} \hlkwc{fit} \hlstd{=} \hlkwa{NULL}\hlstd{,} \hlkwc{scores} \hlstd{= efg5,}
                \hlkwc{order.by} \hlstd{= age,} \hlkwc{parm} \hlstd{=} \hlnum{3}\hlstd{,} \hlkwc{functional} \hlstd{=} \hlstr{'maxLMo'}\hlstd{)}
\end{alltt}
\end{kframe}
\end{knitrout}
\end{figure}

Because this model includes multiple random effects, \pkg{lme4} requires that we use the Laplace approximation (\code{nAGQ = 1}) for estimation. We can still choose a larger number of quadrature points for score computation after model estimation, however, which provides more precise approximations of these quantities. We can also use extra quadrature points to compute the model's log-likelihood (via the \pkg{merDeriv} command \code{llcont.glmerMod()}), which potentially yields a log-likelihood that is more precise than the log-likelihood that is output by \pkg{lme4}.

\subsubsection{Results}
\begin{figure}
\caption{Log-likelihood and standardized gradient of the Poisson mixed model, by number of quadrature points used. The standardized gradient shown is that of the model's fixed intercept parameter.}
\label{fig:poigrad}
\begin{knitrout}
\definecolor{shadecolor}{rgb}{0.969, 0.969, 0.969}\color{fgcolor}
\includegraphics[width=6in,height=3in]{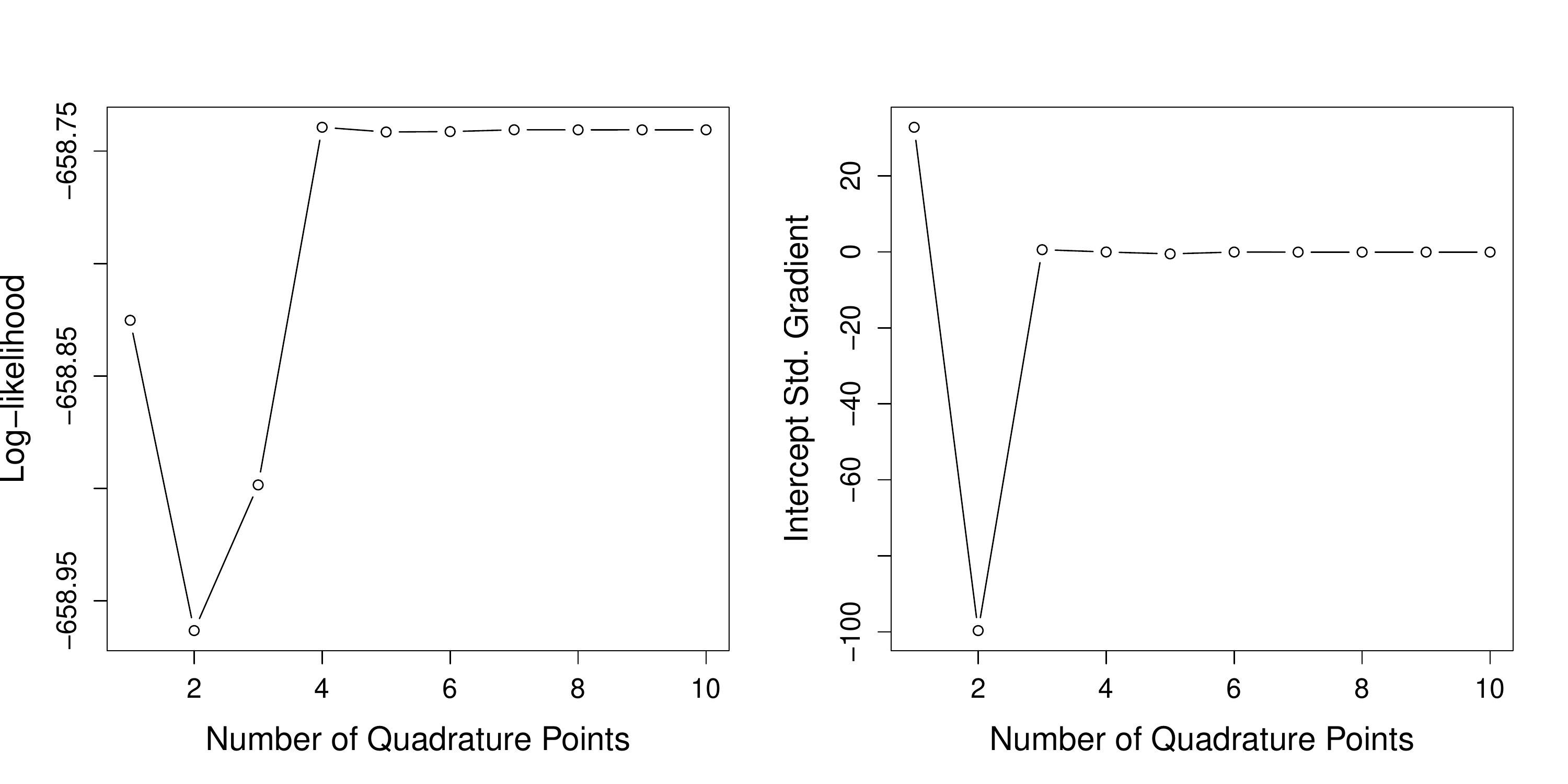} 

\end{knitrout}
\end{figure}

We first used \pkg{merDeriv} to repeatedly compute the log-likelihood and the standardized gradient of the estimated Poisson GLMM, using one to ten quadrature points per dimension (the gradient is obtained by summing scores across people). Some of those results are shown in Figure~\ref{fig:poigrad}, where the left panel displays results for the log-likelihood and the right panel displays results for the standardized gradient of a single model parameter (the fixed intercept). We see that, for small numbers of quadrature points, both of the displayed quantities are unstable. The log-likelihood varies by about a tenth of a point, while the standardized gradient varies by much more. Both quantities stabilize around five quadrature points, however, suggesting that we should use at least that many points in practice (while also considering total computation time).
We also remark that the log-likelihood reported by \pkg{lme4} is the value in the left panel at 1 quadrature point, which is somewhat different from the ``stabilized'' value at larger numbers of quadrature points. We can obtain a more accurate approximation of the fitted model's log-likelihood using the methods described here, and this approximation could influence some likelihood ratio tests or other statistics that rely on the model's log-likelihood.

\begin{table}
  \centering
  \begin{tabular}{lcccccccc} \hline
 & $\beta_0$ & $\beta_1$ & $\beta_2$ & $\beta_3$ & $\beta_4$ & $\beta_5$ & $\beta_6$ & $\beta_7$ \\ \hline
lme4 & 0.1575 & 0.1465 & 0.2194 & 0.0468 & 0.1914 & 0.0404 & 0.0656 & 0.0518 \\
sandwich & 0.2464 & 0.1588 & 0.2301 & 0.0771 & 0.1554 & 0.0452 & 0.0679 & 0.0412 \\ \hline

  \end{tabular}
  \caption{Comparison of model generated standard errors to robust standard errors for Poisson model.}
  \label{tab:poirob}
\end{table}

We now illustrate how methods from the previous sections can be applied to the Poisson GLMM. We first calculate robust standard errors using the code in the middle of Figure~\ref{fig:cc8}, with Table~\ref{tab:poirob} showing the results. The table shows that, for the model considered here, the Huber-White standard errors are generally larger.

\begin{figure}
\caption{Graph of M-Fluctuation test for Poisson model. Model parameters are stable across the range of age.}
\label{fig:poimfluc}
  \begin{center}
\begin{knitrout}
\definecolor{shadecolor}{rgb}{0.969, 0.969, 0.969}\color{fgcolor}
\includegraphics[width=3.5in]{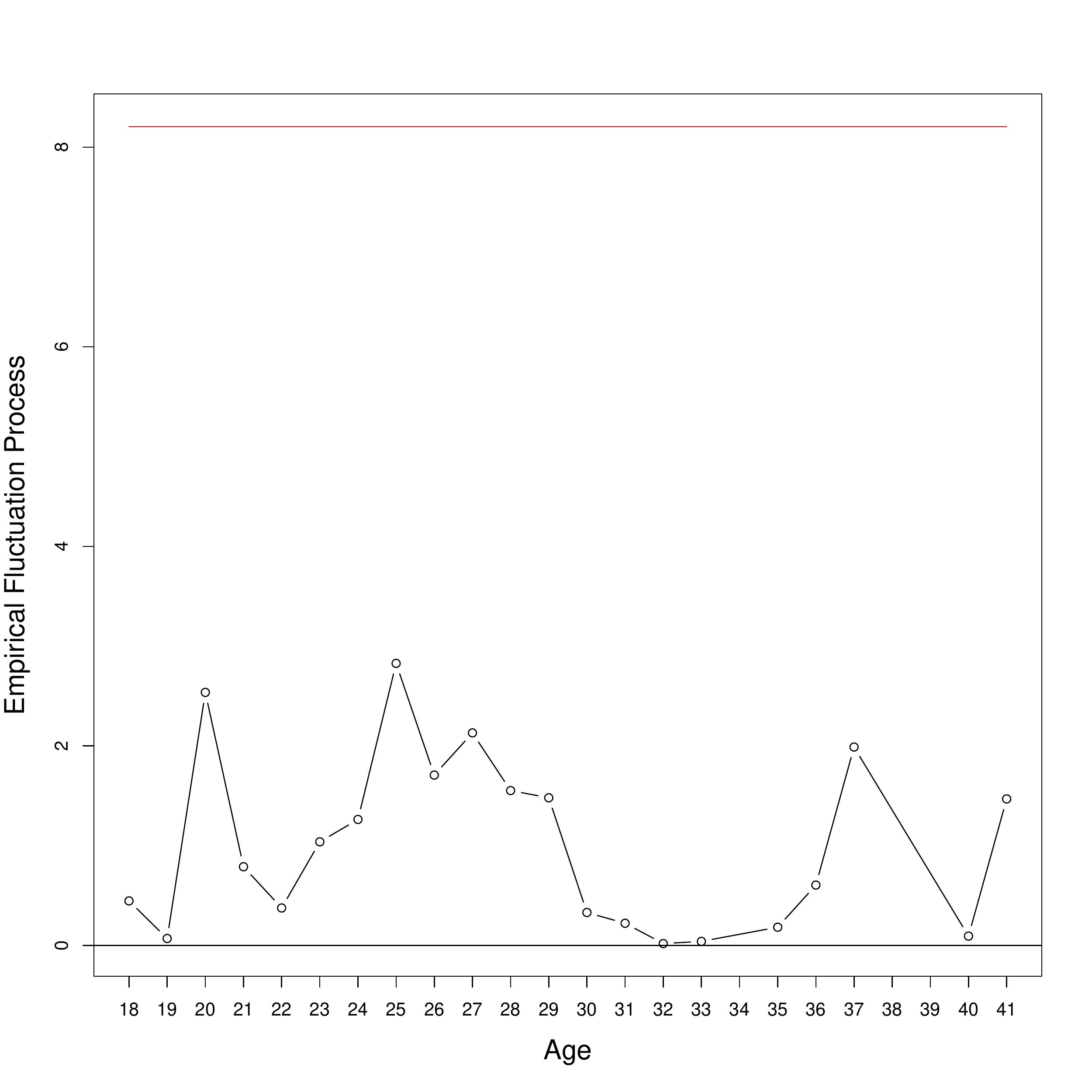} 

\end{knitrout}
\end{center}
\end{figure}

Similarly to the previous section on score-based tests, we next examine the Poisson GLMM parameter fluctuation across an extraneous variable. In this example, we assess the stability of the treatment main effect across patient age, which provides information about whether the treatment efficacy varies for patients of different ages. The score test is carried out via the code at the bottom of Figure~\ref{fig:cc8}, which is similar to that used in the score test section above. The test statistic here (not shown) indicates that the parameter fluctuation is not significant, suggesting that the treatment effect does not fluctuate across the range of age. Figure~\ref{fig:poimfluc} contains the parameter fluctuation across values age, with the critical value being the red horizontal line.

As can be seen, our methods work for other exponential family GLMMs, with the code remaining very similar. In the General Discussion below, we provide further detail about models that our methods cannot handle, as well as future extensions.

\section{General Discussion}
In this paper, we have provided technical details on computing derivatives of the GLMM likelihood function, with a focus on models estimated via package \pkg{lme4}. We then showed how the derivatives can be used in various manners: to obtain robust standard errors, to test predictors that were not included in the estimated model, and to carry out Vuong tests of non-nested GLMMs. All of these applications used the GLMM derivatives in concert with other \proglang{R} packages, illustrating how the \proglang{R} infrastructure can be combined to obtain new statistics that were difficult or impossible to obtain previously.

\subsection{Computational Issues}
The quadrature implementation described in this paper can be uesd to obtain
derivatives of the marginal likelihood function 
for many models with random effects. This method is especially relevant because the conditional random effects $\bm b$ and corresponding components in the variance covariance matrix $\bm G$ are often employed in the model estimation process, in place of derivatives \cite<for example,>{cai2010, cai2010b, bau2004}. Therefore, the derivatives based on the marginal distribution are often not available, or at least not easy to obtain.
Our quadrature method took advantage of the fact that the model was already estimated, so that the predicted modes of the random parameters were available.

Another integral approximation method is the Laplace approximation, which is equivalent to Gauss-Hermite quadrature with one quadrature point \cite{mcc2005}. Thus, the Laplace approximation is less accurate than Gauss-Hermite quadrature with multiple points, but also less computationally intensive and more flexible \cite{stroup12}.
Additionally, it is possible to use derivatives associated with the pseudo maximum likelihood function, which is a transformation of the $y$ response variable into $y^{\star}$, which conditions on the random effect \cite{stroup12}. The scores are then related to a simpler GLM, with such a procedure being implemented in SAS \cite{SAS}. However, the scores based on this pseudo likelihood are not always applicable because the estimates can be problematic, such as when $y$ follows a two-parameter exponential family distribution or sparse Bernoulli distribution \cite{nel92}. 
Finally, numerical methods and Monte Carlo can be flexibly applied to many types of derivative computations, but they are often too slow to be practical.
In all, these remarks indicate that there is not a single, superior method for all scenarios. The quadrature method described here is flexible and appears to work well enough for many types of models.

\subsection{Additional Applications}
There exist other relevant applications that are worth exploring in more detail, including use of the derivatives in GLMM trees. GLMM trees are part of a model-based recursive partitioning framework that has been developed by Zeileis and colleagues \cite{party}. The goal of the framework is to split a dataset into homogeneous subsamples based on auxiliary variables, where each subsample exhibits different values of model parameters. To accomplish this, a tree is constructed via the following steps
\begin{enumerate}
  \item Fit the model of interest to the data in the current node of the tree.
  \item Conduct a score-based test for each auxiliary variable.
  \item Split the current node into two nodes, based on the auxiliary variable with the largest test statistic.
  \item Repeat steps 1--3 for the two nodes that were just created.
\end{enumerate}
This procedure is continued until the score-based tests indicate no parameter instabilities with respect to any auxiliary variables (or until a minimal subsample size is reached).

\citeA{glmertree} recently applied model-based recursive partitioning to GLMMs.  But, due to the difficulty of obtaining scores associated with GLMMs, they developed an alternative procedure where only fixed effect parameters were allowed to vary across subgroups. The developments in the current paper make it possible to apply the original, model-based recursive partitioning procedure to GLMMs, allowing us to detect new types of GLMM heterogeneity in a tree-based framework.

In addition to trees, scores may be used to study heterogeneity through ``on the fly'' tests of residual covariance structures in GLMMs. These developments could reduce computation time by testing multiple covariance structures after fitting a single model, as opposed to requiring estimation of one model per covariance structure. Such tests can be facilitated by the \code{coeftest()} function of package \pkg{lmtest} \cite{zeihot02}, though some \pkg{merDeriv} extensions may be necessary before this works.

\subsection{Limitations}
While the derivations in this paper work for general, exponential family models, two-parameter distributions such as the gamma and inverse Gaussian are additionally complicated by estimation of the extra dispersion parameter. The current \pkg{merDeriv} implementation does not currently handle some of these models, nor does it handle the quasi-Poisson or quasi-binomial families (which are not based on formal likelihood functions). 
Additionally, the applications in this paper take advantage of the fact that we focused on models with a single clustering variable. Researchers often consider three-level models and models with crossed or partially-crossed random effects, though, which utilize multiple clustering variables. The derivations in this paper generally work for those models, allowing us to obtain scores for each case in the data (i.e., for each row of the data). But most of the applications in this paper require a way to split observations into independent groups, which is often impossible when we have multiple clustering variables. For example, individuals in separate groups under one clustering variable may appear in the same group under another clustering variable, leading to different forms of dependence between different pairs of individuals' scores. In contrast, when there is only one clustering variable, we know that individuals in one group are independent of individuals in other groups.

For models with multiple clustering variables, it may be possible to de-correlate scores after the fact, using an appropriately-specified covariance matrix \cite{sand1, sandwichc} or a self-normalization technique that is commonly used in time series research \cite{shao10, zhang11}. This would allow us to split observations into uncorrelated groups, which may be sufficient for applications. Alternatively, \citeA{rasgol94} describe methods for re-specifying a model with crossed random effects to be a fully hierarchical model, in which case it may be possible to directly use the results described in this paper. None of these solutions is trivial, and we hope to further study them in the future. We aspire to a future version of \pkg{merDeriv} that is able to handle all of the models that \pkg{lme4} can estimate.

\section*{Computational Note}
All results were obtained using the \proglang{R}~system for statistical
computing \cite{R20}, version~3.6.1,
employing the add-on package \pkg{merDeriv}~0.2-3 for derivative computations and \pkg{lme4}~1.1-26
\cite{lme4} for fitting of the mixed models. Code to reproduce the results in the paper is available at \url{https://osf.io/58ruw/}.

\bibliography{refs}

\begin{thebibliography}{}

\bibitem [\protect \citeauthoryear {%
Barr%
, Levy%
, Scheepers%
\BCBL {}\ \BBA {} Tily%
}{%
Barr%
\ \protect \BOthers {.}}{%
{\protect \APACyear {2013}}%
}]{%
barlev13}
\APACinsertmetastar {%
barlev13}%
\begin{APACrefauthors}%
Barr, D\BPBI J.%
, Levy, R.%
, Scheepers, C.%
\BCBL {}\ \BBA {} Tily, H\BPBI J.%
\end{APACrefauthors}%
\unskip\
\newblock
\APACrefYearMonthDay{2013}{}{}.
\newblock
{\BBOQ}\APACrefatitle {Random effects structure for confirmatory hypothesis
  testing: {Keep} it maximal} {Random effects structure for confirmatory
  hypothesis testing: {Keep} it maximal}.{\BBCQ}
\newblock
\APACjournalVolNumPages{Journal of Memory and Language}{68}{}{255--278}.
\PrintBackRefs{\CurrentBib}

\bibitem [\protect \citeauthoryear {%
Bates%
}{%
Bates%
}{%
{\protect \APACyear {2021}}%
}]{%
bates2021}
\APACinsertmetastar {%
bates2021}%
\begin{APACrefauthors}%
Bates, D.%
\end{APACrefauthors}%
\unskip\
\newblock
\APACrefYearMonthDay{2021}{}{}.
\newblock
{\BBOQ}\APACrefatitle {Computational methods for mixed models} {Computational
  methods for mixed models}.{\BBCQ}
\newblock
\APACjournalVolNumPages{lme4 Package Vignette}{}{}{}.
\newblock
\begin{APACrefURL}
  \url{https://cran.r-project.org/web/packages/lme4/vignettes/Theory.pdf}
  \end{APACrefURL}
\PrintBackRefs{\CurrentBib}

\bibitem [\protect \citeauthoryear {%
Bates%
, M{\"a}chler%
, Bolker%
\BCBL {}\ \BBA {} Walker%
}{%
Bates%
\ \protect \BOthers {.}}{%
{\protect \APACyear {2015}}%
}]{%
lme4}
\APACinsertmetastar {%
lme4}%
\begin{APACrefauthors}%
Bates, D.%
, M{\"a}chler, M.%
, Bolker, B.%
\BCBL {}\ \BBA {} Walker, S.%
\end{APACrefauthors}%
\unskip\
\newblock
\APACrefYearMonthDay{2015}{}{}.
\newblock
{\BBOQ}\APACrefatitle {Fitting Linear Mixed-Effects Models Using \pkg{lme4}}
  {Fitting linear mixed-effects models using \pkg{lme4}}.{\BBCQ}
\newblock
\APACjournalVolNumPages{Journal of Statistical Software}{67}{1}{1--48}.
\newblock
\begin{APACrefDOI} \doi{10.18637/jss.v067.i01} \end{APACrefDOI}
\PrintBackRefs{\CurrentBib}

\bibitem [\protect \citeauthoryear {%
Bauer%
\ \BBA {} Curran%
}{%
Bauer%
\ \BBA {} Curran%
}{%
{\protect \APACyear {2004}}%
}]{%
bau2004}
\APACinsertmetastar {%
bau2004}%
\begin{APACrefauthors}%
Bauer, D\BPBI J.%
\BCBT {}\ \BBA {} Curran, P\BPBI J.%
\end{APACrefauthors}%
\unskip\
\newblock
\APACrefYearMonthDay{2004}{}{}.
\newblock
{\BBOQ}\APACrefatitle {The integration of continuous and discrete latent
  variable models: {Potential} problems and promising opportunities} {The
  integration of continuous and discrete latent variable models: {Potential}
  problems and promising opportunities}.{\BBCQ}
\newblock
\APACjournalVolNumPages{Psychological Methods}{9}{1}{3--29}.
\PrintBackRefs{\CurrentBib}

\bibitem [\protect \citeauthoryear {%
Bock%
\ \BBA {} Lieberman%
}{%
Bock%
\ \BBA {} Lieberman%
}{%
{\protect \APACyear {1970}}%
}]{%
boclie70}
\APACinsertmetastar {%
boclie70}%
\begin{APACrefauthors}%
Bock, R\BPBI D.%
\BCBT {}\ \BBA {} Lieberman, M.%
\end{APACrefauthors}%
\unskip\
\newblock
\APACrefYearMonthDay{1970}{}{}.
\newblock
{\BBOQ}\APACrefatitle {Fitting a response model for $n$ dichotomously scored
  items} {Fitting a response model for $n$ dichotomously scored items}.{\BBCQ}
\newblock
\APACjournalVolNumPages{Psychometrika}{35}{}{179--198}.
\PrintBackRefs{\CurrentBib}

\bibitem [\protect \citeauthoryear {%
B\"{u}rkner%
}{%
B\"{u}rkner%
}{%
{\protect \APACyear {2018}}%
}]{%
brms}
\APACinsertmetastar {%
brms}%
\begin{APACrefauthors}%
B\"{u}rkner, P\BHBI C.%
\end{APACrefauthors}%
\unskip\
\newblock
\APACrefYearMonthDay{2018}{}{}.
\newblock
{\BBOQ}\APACrefatitle {Advanced {Bayesian} Multilevel Modeling with the {R}
  Package {brms}} {Advanced {Bayesian} multilevel modeling with the {R} package
  {brms}}.{\BBCQ}
\newblock
\APACjournalVolNumPages{The R Journal}{10}{1}{395--411}.
\newblock
\begin{APACrefDOI} \doi{10.32614/RJ-2018-017} \end{APACrefDOI}
\PrintBackRefs{\CurrentBib}

\bibitem [\protect \citeauthoryear {%
Cai%
}{%
Cai%
}{%
{\protect \APACyear {2010}}%
{\protect \APACexlab {{\protect \BCnt {1}}}}}]{%
cai2010b}
\APACinsertmetastar {%
cai2010b}%
\begin{APACrefauthors}%
Cai, L.%
\end{APACrefauthors}%
\unskip\
\newblock
\APACrefYearMonthDay{2010{\protect \BCnt {1}}}{}{}.
\newblock
{\BBOQ}\APACrefatitle {High-dimensional exploratory item factor analysis by a
  {Metropolis-Hastings Robbins-Monro} algorithm} {High-dimensional exploratory
  item factor analysis by a {Metropolis-Hastings Robbins-Monro}
  algorithm}.{\BBCQ}
\newblock
\APACjournalVolNumPages{Psychometrika}{75}{1}{33--57}.
\PrintBackRefs{\CurrentBib}

\bibitem [\protect \citeauthoryear {%
Cai%
}{%
Cai%
}{%
{\protect \APACyear {2010}}%
{\protect \APACexlab {{\protect \BCnt {2}}}}}]{%
cai2010}
\APACinsertmetastar {%
cai2010}%
\begin{APACrefauthors}%
Cai, L.%
\end{APACrefauthors}%
\unskip\
\newblock
\APACrefYearMonthDay{2010{\protect \BCnt {2}}}{}{}.
\newblock
{\BBOQ}\APACrefatitle {A two-tier full-information item factor analysis model
  with applications} {A two-tier full-information item factor analysis model
  with applications}.{\BBCQ}
\newblock
\APACjournalVolNumPages{Psychometrika}{75}{4}{581--612}.
\PrintBackRefs{\CurrentBib}

\bibitem [\protect \citeauthoryear {%
Chalmers%
}{%
Chalmers%
}{%
{\protect \APACyear {2012}}%
}]{%
mirt}
\APACinsertmetastar {%
mirt}%
\begin{APACrefauthors}%
Chalmers, R\BPBI P.%
\end{APACrefauthors}%
\unskip\
\newblock
\APACrefYearMonthDay{2012}{}{}.
\newblock
{\BBOQ}\APACrefatitle {{mirt}: A Multidimensional Item Response Theory Package
  for the \proglang{R} Environment} {{mirt}: A multidimensional item response
  theory package for the \proglang{R} environment}.{\BBCQ}
\newblock
\APACjournalVolNumPages{Journal of Statistical Software}{48}{6}{1--29}.
\newblock
\begin{APACrefDOI} \doi{10.18637/jss.v048.i06} \end{APACrefDOI}
\PrintBackRefs{\CurrentBib}

\bibitem [\protect \citeauthoryear {%
De~Boeck%
\ \protect \BOthers {.}}{%
De~Boeck%
\ \protect \BOthers {.}}{%
{\protect \APACyear {2011}}%
}]{%
de2011}
\APACinsertmetastar {%
de2011}%
\begin{APACrefauthors}%
De~Boeck, P.%
, Bakker, M.%
, Zwitser, R.%
, Nivard, M.%
, Hofman, A.%
, Tuerlinckx, F.%
\BCBL {}\ \BBA {} Partchev, I.%
\end{APACrefauthors}%
\unskip\
\newblock
\APACrefYearMonthDay{2011}{}{}.
\newblock
{\BBOQ}\APACrefatitle {The estimation of item response models with the lmer
  function from the lme4 package in {R}} {The estimation of item response
  models with the lmer function from the lme4 package in {R}}.{\BBCQ}
\newblock
\APACjournalVolNumPages{Journal of Statistical Software}{39}{12}{1--28}.
\PrintBackRefs{\CurrentBib}

\bibitem [\protect \citeauthoryear {%
De~Boeck%
\ \BBA {} Wilson%
}{%
De~Boeck%
\ \BBA {} Wilson%
}{%
{\protect \APACyear {2004}}%
}]{%
debwil04}
\APACinsertmetastar {%
debwil04}%
\begin{APACrefauthors}%
De~Boeck, P.%
\BCBT {}\ \BBA {} Wilson, M.%
\end{APACrefauthors}%
\unskip\
\newblock
\APACrefYear{2004}.
\newblock
\APACrefbtitle {Explanatory item response models: {A} generalized linear and
  nonlinear approach} {Explanatory item response models: {A} generalized linear
  and nonlinear approach}.
\newblock
\APACaddressPublisher{}{New York:\ Springer-Verlag}.
\PrintBackRefs{\CurrentBib}

\bibitem [\protect \citeauthoryear {%
Doran%
, Bates%
, Bliese%
\BCBL {}\ \BBA {} Dowling%
}{%
Doran%
\ \protect \BOthers {.}}{%
{\protect \APACyear {2007}}%
}]{%
dorbat07}
\APACinsertmetastar {%
dorbat07}%
\begin{APACrefauthors}%
Doran, H.%
, Bates, D.%
, Bliese, P.%
\BCBL {}\ \BBA {} Dowling, M.%
\end{APACrefauthors}%
\unskip\
\newblock
\APACrefYearMonthDay{2007}{}{}.
\newblock
{\BBOQ}\APACrefatitle {Estimating the Multilevel {Rasch} Model: {With} the lme4
  Package} {Estimating the multilevel {Rasch} model: {With} the lme4
  package}.{\BBCQ}
\newblock
\APACjournalVolNumPages{Journal of Statistical Software}{20}{2}{1--18}.
\newblock
\begin{APACrefDOI} \doi{10.18637/jss.v020.i02} \end{APACrefDOI}
\PrintBackRefs{\CurrentBib}

\bibitem [\protect \citeauthoryear {%
Embretson%
\ \BBA {} Reise%
}{%
Embretson%
\ \BBA {} Reise%
}{%
{\protect \APACyear {2000}}%
}]{%
embrei00}
\APACinsertmetastar {%
embrei00}%
\begin{APACrefauthors}%
Embretson, S\BPBI E.%
\BCBT {}\ \BBA {} Reise, S\BPBI P.%
\end{APACrefauthors}%
\unskip\
\newblock
\APACrefYear{2000}.
\newblock
\APACrefbtitle {Item response theory for psychologists} {Item response theory
  for psychologists}.
\newblock
\APACaddressPublisher{}{Mahwah, NJ:\ Erlbaum Associates}.
\PrintBackRefs{\CurrentBib}

\bibitem [\protect \citeauthoryear {%
Engle%
}{%
Engle%
}{%
{\protect \APACyear {1984}}%
}]{%
eng84}
\APACinsertmetastar {%
eng84}%
\begin{APACrefauthors}%
Engle, R\BPBI F.%
\end{APACrefauthors}%
\unskip\
\newblock
\APACrefYearMonthDay{1984}{}{}.
\newblock
{\BBOQ}\APACrefatitle {{W}ald, likelihood ratio, and {L}agrange multiplier
  tests in econometrics} {{W}ald, likelihood ratio, and {L}agrange multiplier
  tests in econometrics}.{\BBCQ}
\newblock
\BIn{} Z.~Griliches\ \BBA {} M\BPBI D.~Intriligator\ (\BEDS), \APACrefbtitle
  {Handbook of Econometrics} {Handbook of econometrics}\ (\BVOL~II).
\newblock
\APACaddressPublisher{}{Elsevier}.
\PrintBackRefs{\CurrentBib}

\bibitem [\protect \citeauthoryear {%
Fokkema%
, Smits%
, Zeileis%
, Hothorn%
\BCBL {}\ \BBA {} Kelderman%
}{%
Fokkema%
\ \protect \BOthers {.}}{%
{\protect \APACyear {2018}}%
}]{%
glmertree}
\APACinsertmetastar {%
glmertree}%
\begin{APACrefauthors}%
Fokkema, M.%
, Smits, N.%
, Zeileis, A.%
, Hothorn, T.%
\BCBL {}\ \BBA {} Kelderman, H.%
\end{APACrefauthors}%
\unskip\
\newblock
\APACrefYearMonthDay{2018}{}{}.
\newblock
{\BBOQ}\APACrefatitle {Detecting Treatment-Subgroup Interactions in clustered
  Data With Generalized Linear Mixed-Effects Model Trees} {Detecting
  treatment-subgroup interactions in clustered data with generalized linear
  mixed-effects model trees}.{\BBCQ}
\newblock
\APACjournalVolNumPages{Behavior Research Methods}{50}{}{2016-2034}.
\newblock
\begin{APACrefURL}
  \url{http://link.springer.com/article/10.3758/s13428-017-0971-x}
  \end{APACrefURL}
\PrintBackRefs{\CurrentBib}

\bibitem [\protect \citeauthoryear {%
Glas%
}{%
Glas%
}{%
{\protect \APACyear {1992}}%
}]{%
glas92}
\APACinsertmetastar {%
glas92}%
\begin{APACrefauthors}%
Glas, C\BPBI A\BPBI W.%
\end{APACrefauthors}%
\unskip\
\newblock
\APACrefYearMonthDay{1992}{}{}.
\newblock
{\BBOQ}\APACrefatitle {A {Rasch} model with a multivariate distribution of
  ability} {A {Rasch} model with a multivariate distribution of
  ability}.{\BBCQ}
\newblock
\APACjournalVolNumPages{Objective measurement: Theory into
  practice}{1}{}{236--258}.
\PrintBackRefs{\CurrentBib}

\bibitem [\protect \citeauthoryear {%
Glas%
}{%
Glas%
}{%
{\protect \APACyear {1998}}%
}]{%
glas98}
\APACinsertmetastar {%
glas98}%
\begin{APACrefauthors}%
Glas, C\BPBI A\BPBI W.%
\end{APACrefauthors}%
\unskip\
\newblock
\APACrefYearMonthDay{1998}{}{}.
\newblock
{\BBOQ}\APACrefatitle {Detection of differential item functioning using
  {L}agrange multiplier tests} {Detection of differential item functioning
  using {L}agrange multiplier tests}.{\BBCQ}
\newblock
\APACjournalVolNumPages{Statistica Sinica}{8}{3}{647--667}.
\PrintBackRefs{\CurrentBib}

\bibitem [\protect \citeauthoryear {%
Glas%
}{%
Glas%
}{%
{\protect \APACyear {1999}}%
}]{%
glas99}
\APACinsertmetastar {%
glas99}%
\begin{APACrefauthors}%
Glas, C\BPBI A\BPBI W.%
\end{APACrefauthors}%
\unskip\
\newblock
\APACrefYearMonthDay{1999}{}{}.
\newblock
{\BBOQ}\APACrefatitle {Modification indices for the {2-PL} and the nominal
  response model} {Modification indices for the {2-PL} and the nominal response
  model}.{\BBCQ}
\newblock
\APACjournalVolNumPages{Psychometrika}{64}{}{273--294}.
\PrintBackRefs{\CurrentBib}

\bibitem [\protect \citeauthoryear {%
Hothorn%
\ \BBA {} Zeileis%
}{%
Hothorn%
\ \BBA {} Zeileis%
}{%
{\protect \APACyear {2015}}%
}]{%
party}
\APACinsertmetastar {%
party}%
\begin{APACrefauthors}%
Hothorn, T.%
\BCBT {}\ \BBA {} Zeileis, A.%
\end{APACrefauthors}%
\unskip\
\newblock
\APACrefYearMonthDay{2015}{}{}.
\newblock
{\BBOQ}\APACrefatitle {\pkg{partykit}: A Modular Toolkit for Recursive
  Partytioning in \proglang{R}} {\pkg{partykit}: A modular toolkit for
  recursive partytioning in \proglang{R}}.{\BBCQ}
\newblock
\APACjournalVolNumPages{Journal of Machine Learning Research}{16}{}{3905-3909}.
\newblock
\begin{APACrefURL} \url{http://jmlr.org/papers/v16/hothorn15a.html}
  \end{APACrefURL}
\PrintBackRefs{\CurrentBib}

\bibitem [\protect \citeauthoryear {%
Huber%
}{%
Huber%
}{%
{\protect \APACyear {1967}}%
}]{%
huber67}
\APACinsertmetastar {%
huber67}%
\begin{APACrefauthors}%
Huber, P\BPBI J.%
\end{APACrefauthors}%
\unskip\
\newblock
\APACrefYearMonthDay{1967}{}{}.
\newblock
{\BBOQ}\APACrefatitle {The Behavior of Maximum Likelihood Estimates Under
  Nonstandard Conditions} {The behavior of maximum likelihood estimates under
  nonstandard conditions}.{\BBCQ}
\newblock
\BIn{} \APACrefbtitle {Proceedings of the Fifth {Berkeley} Symposium on
  Mathematical Statistics and Probability} {Proceedings of the fifth {Berkeley}
  symposium on mathematical statistics and probability}\ (\BVOL~1, \BPGS\
  221--233).
\PrintBackRefs{\CurrentBib}

\bibitem [\protect \citeauthoryear {%
Komboz%
, Strobl%
\BCBL {}\ \BBA {} Zeileis%
}{%
Komboz%
\ \protect \BOthers {.}}{%
{\protect \APACyear {2018}}%
}]{%
kom18}
\APACinsertmetastar {%
kom18}%
\begin{APACrefauthors}%
Komboz, B.%
, Strobl, C.%
\BCBL {}\ \BBA {} Zeileis, A.%
\end{APACrefauthors}%
\unskip\
\newblock
\APACrefYearMonthDay{2018}{}{}.
\newblock
{\BBOQ}\APACrefatitle {Tree-based global model tests for polytomous {Rasch}
  models} {Tree-based global model tests for polytomous {Rasch} models}.{\BBCQ}
\newblock
\APACjournalVolNumPages{Educational and Psychological
  Measurement}{78}{1}{128--166}.
\PrintBackRefs{\CurrentBib}

\bibitem [\protect \citeauthoryear {%
Liu%
\ \BBA {} Pierce%
}{%
Liu%
\ \BBA {} Pierce%
}{%
{\protect \APACyear {1994}}%
}]{%
liupie94}
\APACinsertmetastar {%
liupie94}%
\begin{APACrefauthors}%
Liu, Q.%
\BCBT {}\ \BBA {} Pierce, D\BPBI A.%
\end{APACrefauthors}%
\unskip\
\newblock
\APACrefYearMonthDay{1994}{}{}.
\newblock
{\BBOQ}\APACrefatitle {A note on {Gauss-Hermite} quadrature} {A note on
  {Gauss-Hermite} quadrature}.{\BBCQ}
\newblock
\APACjournalVolNumPages{Biometrika}{81}{}{624--629}.
\PrintBackRefs{\CurrentBib}

\bibitem [\protect \citeauthoryear {%
Lord%
\ \BBA {} Novick%
}{%
Lord%
\ \BBA {} Novick%
}{%
{\protect \APACyear {1968}}%
}]{%
lornov68}
\APACinsertmetastar {%
lornov68}%
\begin{APACrefauthors}%
Lord, F\BPBI M.%
\BCBT {}\ \BBA {} Novick, M\BPBI R.%
\end{APACrefauthors}%
\unskip\
\newblock
\APACrefYear{1968}.
\newblock
\APACrefbtitle {Statistical theories of mental test scores} {Statistical
  theories of mental test scores}.
\newblock
\APACaddressPublisher{}{Reading, MA: Addison-Wesley}.
\PrintBackRefs{\CurrentBib}

\bibitem [\protect \citeauthoryear {%
Louis%
}{%
Louis%
}{%
{\protect \APACyear {1982}}%
}]{%
lou82}
\APACinsertmetastar {%
lou82}%
\begin{APACrefauthors}%
Louis, T\BPBI A.%
\end{APACrefauthors}%
\unskip\
\newblock
\APACrefYearMonthDay{1982}{}{}.
\newblock
{\BBOQ}\APACrefatitle {Finding the observed information matrix when using the
  {EM} algorithm} {Finding the observed information matrix when using the {EM}
  algorithm}.{\BBCQ}
\newblock
\APACjournalVolNumPages{Journal of the Royal Statistical Society. Series B
  (Methodological)}{}{}{226--233}.
\PrintBackRefs{\CurrentBib}

\bibitem [\protect \citeauthoryear {%
Matuschek%
, Kliegl%
, Vasishth%
, Baayen%
\BCBL {}\ \BBA {} Bates%
}{%
Matuschek%
\ \protect \BOthers {.}}{%
{\protect \APACyear {2017}}%
}]{%
matkli17}
\APACinsertmetastar {%
matkli17}%
\begin{APACrefauthors}%
Matuschek, H.%
, Kliegl, R.%
, Vasishth, S.%
, Baayen, H.%
\BCBL {}\ \BBA {} Bates, D.%
\end{APACrefauthors}%
\unskip\
\newblock
\APACrefYearMonthDay{2017}{}{}.
\newblock
{\BBOQ}\APACrefatitle {Balancing {Type I} error and power in linear mixed
  models} {Balancing {Type I} error and power in linear mixed models}.{\BBCQ}
\newblock
\APACjournalVolNumPages{Journal of Memory and Language}{94}{}{305--315}.
\PrintBackRefs{\CurrentBib}

\bibitem [\protect \citeauthoryear {%
Mc{C}ullagh%
\ \BBA {} Nelder%
}{%
Mc{C}ullagh%
\ \BBA {} Nelder%
}{%
{\protect \APACyear {1989}}%
}]{%
mccnel89}
\APACinsertmetastar {%
mccnel89}%
\begin{APACrefauthors}%
Mc{C}ullagh, P.%
\BCBT {}\ \BBA {} Nelder, J\BPBI A.%
\end{APACrefauthors}%
\unskip\
\newblock
\APACrefYear{1989}.
\newblock
\APACrefbtitle {Generalized linear models} {Generalized linear models}.
\newblock
\APACaddressPublisher{}{Boca Raton, FL:\ Chapman \& Hall}.
\PrintBackRefs{\CurrentBib}

\bibitem [\protect \citeauthoryear {%
McCulloch%
\ \BBA {} Neuhaus%
}{%
McCulloch%
\ \BBA {} Neuhaus%
}{%
{\protect \APACyear {2001}}%
}]{%
mcc01}
\APACinsertmetastar {%
mcc01}%
\begin{APACrefauthors}%
McCulloch, C\BPBI E.%
\BCBT {}\ \BBA {} Neuhaus, J\BPBI M.%
\end{APACrefauthors}%
\unskip\
\newblock
\APACrefYear{2001}.
\newblock
\APACrefbtitle {Generalized linear mixed models} {Generalized linear mixed
  models}.
\newblock
\APACaddressPublisher{}{Wiley Online Library}.
\PrintBackRefs{\CurrentBib}

\bibitem [\protect \citeauthoryear {%
McCulloch%
\ \BBA {} Neuhaus%
}{%
McCulloch%
\ \BBA {} Neuhaus%
}{%
{\protect \APACyear {2005}}%
}]{%
mcc2005}
\APACinsertmetastar {%
mcc2005}%
\begin{APACrefauthors}%
McCulloch, C\BPBI E.%
\BCBT {}\ \BBA {} Neuhaus, J\BPBI M.%
\end{APACrefauthors}%
\unskip\
\newblock
\APACrefYearMonthDay{2005}{}{}.
\newblock
{\BBOQ}\APACrefatitle {Generalized linear mixed models} {Generalized linear
  mixed models}.{\BBCQ}
\newblock
\APACjournalVolNumPages{Encyclopedia of biostatistics}{4}{}{}.
\PrintBackRefs{\CurrentBib}

\bibitem [\protect \citeauthoryear {%
Merkle%
, Fan%
\BCBL {}\ \BBA {} Zeileis%
}{%
Merkle%
\ \protect \BOthers {.}}{%
{\protect \APACyear {2014}}%
}]{%
merfanzei}
\APACinsertmetastar {%
merfanzei}%
\begin{APACrefauthors}%
Merkle, E\BPBI C.%
, Fan, J.%
\BCBL {}\ \BBA {} Zeileis, A.%
\end{APACrefauthors}%
\unskip\
\newblock
\APACrefYearMonthDay{2014}{}{}.
\newblock
{\BBOQ}\APACrefatitle {Testing for Measurement Invariance with Respect to an
  Ordinal Variable} {Testing for measurement invariance with respect to an
  ordinal variable}.{\BBCQ}
\newblock
\APACjournalVolNumPages{Psychometrika}{79}{}{569--584}.
\PrintBackRefs{\CurrentBib}

\bibitem [\protect \citeauthoryear {%
Merkle%
, Furr%
\BCBL {}\ \BBA {} Rabe-Hesketh%
}{%
Merkle%
\ \protect \BOthers {.}}{%
{\protect \APACyear {2019}}%
}]{%
merfur19}
\APACinsertmetastar {%
merfur19}%
\begin{APACrefauthors}%
Merkle, E\BPBI C.%
, Furr, D.%
\BCBL {}\ \BBA {} Rabe-Hesketh, S.%
\end{APACrefauthors}%
\unskip\
\newblock
\APACrefYearMonthDay{2019}{}{}.
\newblock
{\BBOQ}\APACrefatitle {{Bayesian} comparison of latent variable models:
  {Conditional} versus marginal likelihoods} {{Bayesian} comparison of latent
  variable models: {Conditional} versus marginal likelihoods}.{\BBCQ}
\newblock
\APACjournalVolNumPages{Psychometrika}{84}{}{802--829}.
\PrintBackRefs{\CurrentBib}

\bibitem [\protect \citeauthoryear {%
Merkle%
\ \BBA {} You%
}{%
Merkle%
\ \BBA {} You%
}{%
{\protect \APACyear {2018}}%
}]{%
nonnest2}
\APACinsertmetastar {%
nonnest2}%
\begin{APACrefauthors}%
Merkle, E\BPBI C.%
\BCBT {}\ \BBA {} You, D.%
\end{APACrefauthors}%
\unskip\
\newblock
\APACrefYearMonthDay{2018}{}{}.
\newblock
{\BBOQ}\APACrefatitle {\pkg{nonnest2}: Tests of Non-Nested Models}
  {\pkg{nonnest2}: Tests of non-nested models}{\BBCQ}\
  [\bibcomputersoftwaremanual].
\newblock
\begin{APACrefURL} \url{https://cran.r-project.org/package=nonnest2}
  \end{APACrefURL}
\newblock
\APACrefnote{\proglang{R}~package version~0.5-2}
\PrintBackRefs{\CurrentBib}

\bibitem [\protect \citeauthoryear {%
Merkle%
, You%
\BCBL {}\ \BBA {} Preacher%
}{%
Merkle%
\ \protect \BOthers {.}}{%
{\protect \APACyear {2016}}%
}]{%
merkle16}
\APACinsertmetastar {%
merkle16}%
\begin{APACrefauthors}%
Merkle, E\BPBI C.%
, You, D.%
\BCBL {}\ \BBA {} Preacher, K\BPBI J.%
\end{APACrefauthors}%
\unskip\
\newblock
\APACrefYearMonthDay{2016}{}{}.
\newblock
{\BBOQ}\APACrefatitle {Testing Nonnested Structural Equation Models.} {Testing
  nonnested structural equation models.}{\BBCQ}
\newblock
\APACjournalVolNumPages{Psychological Methods}{21}{2}{151--163}.
\PrintBackRefs{\CurrentBib}

\bibitem [\protect \citeauthoryear {%
Merkle%
\ \BBA {} Zeileis%
}{%
Merkle%
\ \BBA {} Zeileis%
}{%
{\protect \APACyear {2013}}%
}]{%
merzei13}
\APACinsertmetastar {%
merzei13}%
\begin{APACrefauthors}%
Merkle, E\BPBI C.%
\BCBT {}\ \BBA {} Zeileis, A.%
\end{APACrefauthors}%
\unskip\
\newblock
\APACrefYearMonthDay{2013}{}{}.
\newblock
{\BBOQ}\APACrefatitle {Tests of Measurement Invariance without Subgroups: {A}
  Generalization of Classical Methods} {Tests of measurement invariance without
  subgroups: {A} generalization of classical methods}.{\BBCQ}
\newblock
\APACjournalVolNumPages{Psychometrika}{78}{}{59--82}.
\PrintBackRefs{\CurrentBib}

\bibitem [\protect \citeauthoryear {%
Naylor%
\ \BBA {} Smith%
}{%
Naylor%
\ \BBA {} Smith%
}{%
{\protect \APACyear {1982}}%
}]{%
naysmi82}
\APACinsertmetastar {%
naysmi82}%
\begin{APACrefauthors}%
Naylor, J\BPBI C.%
\BCBT {}\ \BBA {} Smith, A\BPBI F\BPBI M.%
\end{APACrefauthors}%
\unskip\
\newblock
\APACrefYearMonthDay{1982}{}{}.
\newblock
{\BBOQ}\APACrefatitle {Applications of a method for the efficient computation
  of posterior distributions} {Applications of a method for the efficient
  computation of posterior distributions}.{\BBCQ}
\newblock
\APACjournalVolNumPages{Journal of the Royal Statistical Society
  C}{31}{}{214--225}.
\PrintBackRefs{\CurrentBib}

\bibitem [\protect \citeauthoryear {%
Nelder%
\ \BBA {} Lee%
}{%
Nelder%
\ \BBA {} Lee%
}{%
{\protect \APACyear {1992}}%
}]{%
nel92}
\APACinsertmetastar {%
nel92}%
\begin{APACrefauthors}%
Nelder, J.%
\BCBT {}\ \BBA {} Lee, Y.%
\end{APACrefauthors}%
\unskip\
\newblock
\APACrefYearMonthDay{1992}{}{}.
\newblock
{\BBOQ}\APACrefatitle {Likelihood, quasi-likelihood and pseudolikelihood: some
  comparisons} {Likelihood, quasi-likelihood and pseudolikelihood: some
  comparisons}.{\BBCQ}
\newblock
\APACjournalVolNumPages{Journal of the Royal Statistical Society: Series B
  (Methodological)}{54}{1}{273--284}.
\PrintBackRefs{\CurrentBib}

\bibitem [\protect \citeauthoryear {%
{Open Source Psychometrics Project}%
}{%
{Open Source Psychometrics Project}%
}{%
{\protect \APACyear {{\protect \bibnodate {}}}}%
}]{%
ospp}
\APACinsertmetastar {%
ospp}%
\begin{APACrefauthors}%
{Open Source Psychometrics Project}.%
\end{APACrefauthors}%
\unskip\
\newblock
\APACrefYearMonthDay{{\protect \bibnodate {}}}{}{}.
\newblock
{\BBOQ}\APACrefatitle {Open psychology data: Raw data from online} {Open
  psychology data: Raw data from online}{\BBCQ}\ [\bibcomputersoftwaremanual].
\newblock
\begin{APACrefURL} [{2017-10-17}]\url{https://openpsychometrics.org/_rawdata/}
  \end{APACrefURL}
\PrintBackRefs{\CurrentBib}

\bibitem [\protect \citeauthoryear {%
Petersen%
\ \BBA {} Pedersen%
}{%
Petersen%
\ \BBA {} Pedersen%
}{%
{\protect \APACyear {2012}}%
}]{%
peter08}
\APACinsertmetastar {%
peter08}%
\begin{APACrefauthors}%
Petersen, K\BPBI B.%
\BCBT {}\ \BBA {} Pedersen, M\BPBI S.%
\end{APACrefauthors}%
\unskip\
\newblock
\APACrefYearMonthDay{2012}{}{}.
\newblock
\APACrefbtitle {The Matrix Cookbook.} {The matrix cookbook.}
\newblock
\APACaddressPublisher{}{Technical University of Denmark}.
\newblock
\begin{APACrefURL} \url{http://www2.imm.dtu.dk/pubdb/p.php?3274}
  \end{APACrefURL}
\newblock
\APACrefnote{Version 20121115}
\PrintBackRefs{\CurrentBib}

\bibitem [\protect \citeauthoryear {%
Pinheiro%
\ \BBA {} Bates%
}{%
Pinheiro%
\ \BBA {} Bates%
}{%
{\protect \APACyear {1995}}%
}]{%
pinbat95}
\APACinsertmetastar {%
pinbat95}%
\begin{APACrefauthors}%
Pinheiro, J\BPBI C.%
\BCBT {}\ \BBA {} Bates, D\BPBI M.%
\end{APACrefauthors}%
\unskip\
\newblock
\APACrefYearMonthDay{1995}{}{}.
\newblock
{\BBOQ}\APACrefatitle {Approximations to the log-likelihood function in the
  nonlinear mixed-effects model} {Approximations to the log-likelihood function
  in the nonlinear mixed-effects model}.{\BBCQ}
\newblock
\APACjournalVolNumPages{Journal of Computational Graphics and
  Statistics}{4}{}{12-35}.
\PrintBackRefs{\CurrentBib}

\bibitem [\protect \citeauthoryear {%
Powell%
}{%
Powell%
}{%
{\protect \APACyear {2009}}%
}]{%
powell2009}
\APACinsertmetastar {%
powell2009}%
\begin{APACrefauthors}%
Powell, M\BPBI J.%
\end{APACrefauthors}%
\unskip\
\newblock
\APACrefYearMonthDay{2009}{}{}.
\newblock
{\BBOQ}\APACrefatitle {The {BOBYQA} algorithm for bound constrained
  optimization without derivatives} {The {BOBYQA} algorithm for bound
  constrained optimization without derivatives}.{\BBCQ}
\newblock
\APACjournalVolNumPages{Cambridge NA Report NA2009/06, University of Cambridge,
  Cambridge}{}{}{26--46}.
\PrintBackRefs{\CurrentBib}

\bibitem [\protect \citeauthoryear {%
{R Core Team}%
}{%
{R Core Team}%
}{%
{\protect \APACyear {2020}}%
}]{%
R20}
\APACinsertmetastar {%
R20}%
\begin{APACrefauthors}%
{R Core Team}.%
\end{APACrefauthors}%
\unskip\
\newblock
\APACrefYearMonthDay{2020}{}{}.
\newblock
{\BBOQ}\APACrefatitle {R: A Language and Environment for Statistical Computing}
  {R: A language and environment for statistical computing}{\BBCQ}\
  [\bibcomputersoftwaremanual].
\newblock
\APACaddressPublisher{Vienna, Austria}{}.
\newblock
\begin{APACrefURL} \url{https://www.R-project.org/} \end{APACrefURL}
\PrintBackRefs{\CurrentBib}

\bibitem [\protect \citeauthoryear {%
Rabe-Hesketh%
, Skrondal%
\BCBL {}\ \BBA {} Pickles%
}{%
Rabe-Hesketh%
\ \protect \BOthers {.}}{%
{\protect \APACyear {2005}}%
}]{%
rab05}
\APACinsertmetastar {%
rab05}%
\begin{APACrefauthors}%
Rabe-Hesketh, S.%
, Skrondal, A.%
\BCBL {}\ \BBA {} Pickles, A.%
\end{APACrefauthors}%
\unskip\
\newblock
\APACrefYearMonthDay{2005}{}{}.
\newblock
{\BBOQ}\APACrefatitle {Maximum likelihood estimation of limited and discrete
  dependent variable models with nested random effects} {Maximum likelihood
  estimation of limited and discrete dependent variable models with nested
  random effects}.{\BBCQ}
\newblock
\APACjournalVolNumPages{Journal of Econometrics}{128}{2}{301--323}.
\PrintBackRefs{\CurrentBib}

\bibitem [\protect \citeauthoryear {%
Rasbash%
\ \BBA {} Goldstein%
}{%
Rasbash%
\ \BBA {} Goldstein%
}{%
{\protect \APACyear {1994}}%
}]{%
rasgol94}
\APACinsertmetastar {%
rasgol94}%
\begin{APACrefauthors}%
Rasbash, J.%
\BCBT {}\ \BBA {} Goldstein, H.%
\end{APACrefauthors}%
\unskip\
\newblock
\APACrefYearMonthDay{1994}{}{}.
\newblock
{\BBOQ}\APACrefatitle {Efficient analysis of mixed hierarchical and
  cross-classified random structures using a multilevel model} {Efficient
  analysis of mixed hierarchical and cross-classified random structures using a
  multilevel model}.{\BBCQ}
\newblock
\APACjournalVolNumPages{Journal of Educational and Behavioral
  Statistics}{19}{}{337--350}.
\PrintBackRefs{\CurrentBib}

\bibitem [\protect \citeauthoryear {%
Schabenberger%
}{%
Schabenberger%
}{%
{\protect \APACyear {2005}}%
}]{%
SAS}
\APACinsertmetastar {%
SAS}%
\begin{APACrefauthors}%
Schabenberger, O.%
\end{APACrefauthors}%
\unskip\
\newblock
\APACrefYearMonthDay{2005}{}{}.
\newblock
{\BBOQ}\APACrefatitle {Introducing the {GLIMMIX} procedure for generalized
  linear mixed models} {Introducing the {GLIMMIX} procedure for generalized
  linear mixed models}.{\BBCQ}
\newblock
\APACjournalVolNumPages{SUGI 30 Proceedings}{196}{}{}.
\PrintBackRefs{\CurrentBib}

\bibitem [\protect \citeauthoryear {%
Schneider%
, Chalmers%
, Debelak%
\BCBL {}\ \BBA {} Merkle%
}{%
Schneider%
\ \protect \BOthers {.}}{%
{\protect \APACyear {2020}}%
}]{%
schcha19}
\APACinsertmetastar {%
schcha19}%
\begin{APACrefauthors}%
Schneider, L.%
, Chalmers, R\BPBI P.%
, Debelak, R.%
\BCBL {}\ \BBA {} Merkle, E\BPBI C.%
\end{APACrefauthors}%
\unskip\
\newblock
\APACrefYearMonthDay{2020}{}{}.
\newblock
{\BBOQ}\APACrefatitle {Model selection of nested and non-nested item response
  models using {Vuong} tests} {Model selection of nested and non-nested item
  response models using {Vuong} tests}.{\BBCQ}
\newblock
\APACjournalVolNumPages{Multivariate Behavioral Research}{55}{}{664--684}.
\PrintBackRefs{\CurrentBib}

\bibitem [\protect \citeauthoryear {%
Shao%
\ \BBA {} Zhang%
}{%
Shao%
\ \BBA {} Zhang%
}{%
{\protect \APACyear {2010}}%
}]{%
shao10}
\APACinsertmetastar {%
shao10}%
\begin{APACrefauthors}%
Shao, X.%
\BCBT {}\ \BBA {} Zhang, X.%
\end{APACrefauthors}%
\unskip\
\newblock
\APACrefYearMonthDay{2010}{}{}.
\newblock
{\BBOQ}\APACrefatitle {Testing for change points in time series} {Testing for
  change points in time series}.{\BBCQ}
\newblock
\APACjournalVolNumPages{Journal of the American Statistical
  Association}{105}{491}{1228--1240}.
\PrintBackRefs{\CurrentBib}

\bibitem [\protect \citeauthoryear {%
Skrondal%
\ \BBA {} Rabe-Hesketh%
}{%
Skrondal%
\ \BBA {} Rabe-Hesketh%
}{%
{\protect \APACyear {2004}}%
}]{%
skrrab04}
\APACinsertmetastar {%
skrrab04}%
\begin{APACrefauthors}%
Skrondal, A.%
\BCBT {}\ \BBA {} Rabe-Hesketh, S.%
\end{APACrefauthors}%
\unskip\
\newblock
\APACrefYear{2004}.
\newblock
\APACrefbtitle {Generalized latent variable modeling:\ {Multilevel,}
  longitudinal, and structural equation modeling} {Generalized latent variable
  modeling:\ {Multilevel,} longitudinal, and structural equation modeling}.
\newblock
\APACaddressPublisher{}{Boca Raton, FL:\ Chapman \& Hall}.
\PrintBackRefs{\CurrentBib}

\bibitem [\protect \citeauthoryear {%
Strobl%
, Kopf%
\BCBL {}\ \BBA {} Zeileis%
}{%
Strobl%
\ \protect \BOthers {.}}{%
{\protect \APACyear {2015}}%
}]{%
strkop15}
\APACinsertmetastar {%
strkop15}%
\begin{APACrefauthors}%
Strobl, C.%
, Kopf, J.%
\BCBL {}\ \BBA {} Zeileis, A.%
\end{APACrefauthors}%
\unskip\
\newblock
\APACrefYearMonthDay{2015}{}{}.
\newblock
{\BBOQ}\APACrefatitle {Rasch Trees: A New Method for Detecting Differential
  Item Functioning in the {R}asch Model} {Rasch trees: A new method for
  detecting differential item functioning in the {R}asch model}.{\BBCQ}
\newblock
\APACjournalVolNumPages{Psychometrika}{80}{2}{289--316}.
\newblock
\begin{APACrefDOI} \doi{10.1007/s11336-013-9388-3} \end{APACrefDOI}
\PrintBackRefs{\CurrentBib}

\bibitem [\protect \citeauthoryear {%
Stroup%
}{%
Stroup%
}{%
{\protect \APACyear {2012}}%
}]{%
stroup12}
\APACinsertmetastar {%
stroup12}%
\begin{APACrefauthors}%
Stroup, W\BPBI W.%
\end{APACrefauthors}%
\unskip\
\newblock
\APACrefYear{2012}.
\newblock
\APACrefbtitle {Generalized Linear Mixed Models: Modern Concepts, Methods and
  Applications} {Generalized linear mixed models: Modern concepts, methods and
  applications}.
\newblock
\APACaddressPublisher{}{CRC Press}.
\PrintBackRefs{\CurrentBib}

\bibitem [\protect \citeauthoryear {%
Stroup%
\ \BBA {} Claassen%
}{%
Stroup%
\ \BBA {} Claassen%
}{%
{\protect \APACyear {2020}}%
}]{%
str20}
\APACinsertmetastar {%
str20}%
\begin{APACrefauthors}%
Stroup, W\BPBI W.%
\BCBT {}\ \BBA {} Claassen, E.%
\end{APACrefauthors}%
\unskip\
\newblock
\APACrefYearMonthDay{2020}{}{}.
\newblock
{\BBOQ}\APACrefatitle {Pseudo-Likelihood or Quadrature? {What} We Thought We
  Knew, What We Think We Know, and What We Are Still Trying to Figure Out}
  {Pseudo-likelihood or quadrature? {What} we thought we knew, what we think we
  know, and what we are still trying to figure out}.{\BBCQ}
\newblock
\APACjournalVolNumPages{Journal of Agricultural, Biological and Environmental
  Statistics}{}{}{}.
\PrintBackRefs{\CurrentBib}

\bibitem [\protect \citeauthoryear {%
Thall%
\ \BBA {} Vail%
}{%
Thall%
\ \BBA {} Vail%
}{%
{\protect \APACyear {1990}}%
}]{%
thall1990}
\APACinsertmetastar {%
thall1990}%
\begin{APACrefauthors}%
Thall, P\BPBI F.%
\BCBT {}\ \BBA {} Vail, S\BPBI C.%
\end{APACrefauthors}%
\unskip\
\newblock
\APACrefYearMonthDay{1990}{}{}.
\newblock
{\BBOQ}\APACrefatitle {Some covariance models for longitudinal count data with
  overdispersion} {Some covariance models for longitudinal count data with
  overdispersion}.{\BBCQ}
\newblock
\APACjournalVolNumPages{Biometrics}{}{}{657--671}.
\PrintBackRefs{\CurrentBib}

\bibitem [\protect \citeauthoryear {%
Trepte%
\ \BBA {} Verbeet%
}{%
Trepte%
\ \BBA {} Verbeet%
}{%
{\protect \APACyear {2010}}%
}]{%
trever10}
\APACinsertmetastar {%
trever10}%
\begin{APACrefauthors}%
Trepte, S.%
\BCBT {}\ \BBA {} Verbeet, M.%
\end{APACrefauthors}%
\ (\BEDS).
\unskip\
\newblock
\APACrefYear{2010}.
\newblock
\APACrefbtitle {{A}llgemeinbildung in {D}eutschland -- Erkenntnisse aus dem
  {SPIEGEL} {S}tudentenpisa-{T}est} {{A}llgemeinbildung in {D}eutschland --
  erkenntnisse aus dem {SPIEGEL} {S}tudentenpisa-{T}est}.
\newblock
\APACaddressPublisher{Wiesbaden}{VS Verlag}.
\PrintBackRefs{\CurrentBib}

\bibitem [\protect \citeauthoryear {%
Vuong%
}{%
Vuong%
}{%
{\protect \APACyear {1989}}%
}]{%
vuo89}
\APACinsertmetastar {%
vuo89}%
\begin{APACrefauthors}%
Vuong, Q\BPBI H.%
\end{APACrefauthors}%
\unskip\
\newblock
\APACrefYearMonthDay{1989}{}{}.
\newblock
{\BBOQ}\APACrefatitle {Likelihood Ratio Tests for Model Selection and
  Non-Nested Hypotheses} {Likelihood ratio tests for model selection and
  non-nested hypotheses}.{\BBCQ}
\newblock
\APACjournalVolNumPages{Econometrica}{57}{}{307--333}.
\PrintBackRefs{\CurrentBib}

\bibitem [\protect \citeauthoryear {%
Wang%
\ \BBA {} Merkle%
}{%
Wang%
\ \BBA {} Merkle%
}{%
{\protect \APACyear {2018}}%
}]{%
merDeriv}
\APACinsertmetastar {%
merDeriv}%
\begin{APACrefauthors}%
Wang, T.%
\BCBT {}\ \BBA {} Merkle, E\BPBI C.%
\end{APACrefauthors}%
\unskip\
\newblock
\APACrefYearMonthDay{2018}{}{}.
\newblock
{\BBOQ}\APACrefatitle {{merDeriv}: Derivative Computations for Linear Mixed
  Effects Models with Application to Robust Standard Errors} {{merDeriv}:
  Derivative computations for linear mixed effects models with application to
  robust standard errors}.{\BBCQ}
\newblock
\APACjournalVolNumPages{Journal of Statistical Software}{87}{1}{1--16}.
\newblock
\begin{APACrefDOI} \doi{10.18637/jss.v087.c01} \end{APACrefDOI}
\PrintBackRefs{\CurrentBib}

\bibitem [\protect \citeauthoryear {%
Wang%
, Merkle%
, Anguera%
\BCBL {}\ \BBA {} Turner%
}{%
Wang%
\ \protect \BOthers {.}}{%
{\protect \APACyear {2020}}%
}]{%
wanmer20}
\APACinsertmetastar {%
wanmer20}%
\begin{APACrefauthors}%
Wang, T.%
, Merkle, E\BPBI C.%
, Anguera, J\BPBI A.%
\BCBL {}\ \BBA {} Turner, B\BPBI M.%
\end{APACrefauthors}%
\unskip\
\newblock
\APACrefYearMonthDay{2020}{}{}.
\newblock
{\BBOQ}\APACrefatitle {Score-based tests for detecting heterogeneity in linear
  mixed models} {Score-based tests for detecting heterogeneity in linear mixed
  models}.{\BBCQ}
\newblock
\APACjournalVolNumPages{Behavior Research Methods}{}{}{}.
\PrintBackRefs{\CurrentBib}

\bibitem [\protect \citeauthoryear {%
Wang%
, Strobl%
, Zeileis%
\BCBL {}\ \BBA {} Merkle%
}{%
Wang%
\ \protect \BOthers {.}}{%
{\protect \APACyear {2018}}%
}]{%
wanstr18}
\APACinsertmetastar {%
wanstr18}%
\begin{APACrefauthors}%
Wang, T.%
, Strobl, C.%
, Zeileis, A.%
\BCBL {}\ \BBA {} Merkle, E\BPBI C.%
\end{APACrefauthors}%
\unskip\
\newblock
\APACrefYearMonthDay{2018}{}{}.
\newblock
{\BBOQ}\APACrefatitle {Score-based tests of differential item functioning via
  pairwise maximum likelihood estimation} {Score-based tests of differential
  item functioning via pairwise maximum likelihood estimation}.{\BBCQ}
\newblock
\APACjournalVolNumPages{Psychometrika}{83}{1}{132--155}.
\PrintBackRefs{\CurrentBib}

\bibitem [\protect \citeauthoryear {%
White%
}{%
White%
}{%
{\protect \APACyear {1980}}%
}]{%
white80}
\APACinsertmetastar {%
white80}%
\begin{APACrefauthors}%
White, H.%
\end{APACrefauthors}%
\unskip\
\newblock
\APACrefYearMonthDay{1980}{}{}.
\newblock
{\BBOQ}\APACrefatitle {A Heteroskedasticity-Consistent Covariance Matrix
  Estimator and a Direct Test for Heteroskedasticity} {A
  heteroskedasticity-consistent covariance matrix estimator and a direct test
  for heteroskedasticity}.{\BBCQ}
\newblock
\APACjournalVolNumPages{Econometrica: Journal of the Econometric
  Society}{48}{4}{817--838}.
\PrintBackRefs{\CurrentBib}

\bibitem [\protect \citeauthoryear {%
Zeileis%
}{%
Zeileis%
}{%
{\protect \APACyear {2004}}%
}]{%
sand1}
\APACinsertmetastar {%
sand1}%
\begin{APACrefauthors}%
Zeileis, A.%
\end{APACrefauthors}%
\unskip\
\newblock
\APACrefYearMonthDay{2004}{}{}.
\newblock
{\BBOQ}\APACrefatitle {Econometric Computing with {HC} and {HAC} Covariance
  Matrix Estimators} {Econometric computing with {HC} and {HAC} covariance
  matrix estimators}.{\BBCQ}
\newblock
\APACjournalVolNumPages{Journal of Statistical Software}{11}{10}{1--17}.
\newblock
\begin{APACrefURL} \url{http://www.jstatsoft.org/v11/i10/} \end{APACrefURL}
\PrintBackRefs{\CurrentBib}

\bibitem [\protect \citeauthoryear {%
Zeileis%
}{%
Zeileis%
}{%
{\protect \APACyear {2006}}%
}]{%
sandwichb}
\APACinsertmetastar {%
sandwichb}%
\begin{APACrefauthors}%
Zeileis, A.%
\end{APACrefauthors}%
\unskip\
\newblock
\APACrefYearMonthDay{2006}{}{}.
\newblock
{\BBOQ}\APACrefatitle {Object-Oriented Computation of Sandwich Estimators}
  {Object-oriented computation of sandwich estimators}.{\BBCQ}
\newblock
\APACjournalVolNumPages{Journal of Statistical Software}{16}{9}{1--16}.
\newblock
\begin{APACrefDOI} \doi{10.18637/jss.v016.i09} \end{APACrefDOI}
\PrintBackRefs{\CurrentBib}

\bibitem [\protect \citeauthoryear {%
Zeileis%
\ \BBA {} Hornik%
}{%
Zeileis%
\ \BBA {} Hornik%
}{%
{\protect \APACyear {2007}}%
}]{%
zeihor07}
\APACinsertmetastar {%
zeihor07}%
\begin{APACrefauthors}%
Zeileis, A.%
\BCBT {}\ \BBA {} Hornik, K.%
\end{APACrefauthors}%
\unskip\
\newblock
\APACrefYearMonthDay{2007}{}{}.
\newblock
{\BBOQ}\APACrefatitle {Generalized {M}-Fluctuation Tests for Parameter
  Instability} {Generalized {M}-fluctuation tests for parameter
  instability}.{\BBCQ}
\newblock
\APACjournalVolNumPages{Statistica Neerlandica}{61}{}{488--508}.
\PrintBackRefs{\CurrentBib}

\bibitem [\protect \citeauthoryear {%
Zeileis%
\ \BBA {} Hothorn%
}{%
Zeileis%
\ \BBA {} Hothorn%
}{%
{\protect \APACyear {2002}}%
}]{%
zeihot02}
\APACinsertmetastar {%
zeihot02}%
\begin{APACrefauthors}%
Zeileis, A.%
\BCBT {}\ \BBA {} Hothorn, T.%
\end{APACrefauthors}%
\unskip\
\newblock
\APACrefYearMonthDay{2002}{}{}.
\newblock
{\BBOQ}\APACrefatitle {Diagnostic Checking in Regression Relationships}
  {Diagnostic checking in regression relationships}.{\BBCQ}
\newblock
\APACjournalVolNumPages{\proglang{R} News}{2}{3}{7--10}.
\newblock
\begin{APACrefURL} \url{https://CRAN.R-project.org/doc/Rnews/} \end{APACrefURL}
\PrintBackRefs{\CurrentBib}

\bibitem [\protect \citeauthoryear {%
Zeileis%
, K\"{o}ll%
\BCBL {}\ \BBA {} Graham%
}{%
Zeileis%
\ \protect \BOthers {.}}{%
{\protect \APACyear {2020}}%
}]{%
sandwichc}
\APACinsertmetastar {%
sandwichc}%
\begin{APACrefauthors}%
Zeileis, A.%
, K\"{o}ll, S.%
\BCBL {}\ \BBA {} Graham, N.%
\end{APACrefauthors}%
\unskip\
\newblock
\APACrefYearMonthDay{2020}{}{}.
\newblock
{\BBOQ}\APACrefatitle {Various Versatile Variances: An Object-Oriented
  Implementation of Clustered Covariances in {R}} {Various versatile variances:
  An object-oriented implementation of clustered covariances in {R}}.{\BBCQ}
\newblock
\APACjournalVolNumPages{Journal of Statistical Software}{95}{1}{}.
\newblock
\begin{APACrefDOI} \doi{10.18637/jss.v095.i01} \end{APACrefDOI}
\PrintBackRefs{\CurrentBib}

\bibitem [\protect \citeauthoryear {%
Zeileis%
, Leisch%
, Hornik%
\BCBL {}\ \BBA {} Kleiber%
}{%
Zeileis%
\ \protect \BOthers {.}}{%
{\protect \APACyear {2002}}%
}]{%
strucchange}
\APACinsertmetastar {%
strucchange}%
\begin{APACrefauthors}%
Zeileis, A.%
, Leisch, F.%
, Hornik, K.%
\BCBL {}\ \BBA {} Kleiber, C.%
\end{APACrefauthors}%
\unskip\
\newblock
\APACrefYearMonthDay{2002}{}{}.
\newblock
{\BBOQ}\APACrefatitle {strucchange: {An R} Package for Testing for Structural
  Change in Linear Regression Models} {strucchange: {An R} package for testing
  for structural change in linear regression models}.{\BBCQ}
\newblock
\APACjournalVolNumPages{Journal of Statistical Software}{7}{2}{1--38}.
\newblock
\begin{APACrefURL} \url{http://www.jstatsoft.org/v07/i02/} \end{APACrefURL}
\PrintBackRefs{\CurrentBib}

\bibitem [\protect \citeauthoryear {%
Zhang%
, Shao%
, Hayhoe%
\BCBL {}\ \BBA {} Wuebbles%
}{%
Zhang%
\ \protect \BOthers {.}}{%
{\protect \APACyear {2011}}%
}]{%
zhang11}
\APACinsertmetastar {%
zhang11}%
\begin{APACrefauthors}%
Zhang, X.%
, Shao, X.%
, Hayhoe, K.%
\BCBL {}\ \BBA {} Wuebbles, D\BPBI J.%
\end{APACrefauthors}%
\unskip\
\newblock
\APACrefYearMonthDay{2011}{}{}.
\newblock
{\BBOQ}\APACrefatitle {Testing the structural stability of temporally dependent
  functional observations and application to climate projections} {Testing the
  structural stability of temporally dependent functional observations and
  application to climate projections}.{\BBCQ}
\newblock
\APACjournalVolNumPages{Electronic Journal of Statistics}{5}{}{1765--1796}.
\PrintBackRefs{\CurrentBib}

\end{thebibliography}





\end{document}